\documentclass[lettersize,journal]{IEEEtran}
\usepackage{amsmath,amsfonts}
\usepackage{algorithmic}
\usepackage{array}
\usepackage[caption=false,font=normalsize,labelfont=sf,textfont=sf]{subfig}
\usepackage{textcomp}
\usepackage{stfloats}
\usepackage{url}
\usepackage{verbatim}
\usepackage{graphicx}
\usepackage{multirow}
\usepackage{multicol}
\usepackage{adjustbox}

\def\BibTeX{{\rm B\kern-.05em{\sc i\kern-.025em b}\kern-.08em
    T\kern-.1667em\lower.7ex\hbox{E}\kern-.125emX}}
\usepackage{balance}
\begin{document}
\title{FgC2F-UDiff: Frequency-guided and Coarse-to-fine Unified Diffusion Model for Multi-modality Missing MRI Synthesis}

\author{Xiaojiao Xiao, Qinmin Vivian Hu, and Guanghui Wang, \IEEEmembership{Senior Member, IEEE}
\thanks{This work is partly supported by the Natural Sciences and Engineering Research Council of Canada (NSERC) and TMU FOS Postdoctoral Fellowship.}
\thanks{X. Xiao and Q. Hu are with the Department of Computer Science, Toronto Metropolitan University, Toronto, Canada (e-mail:xiaojiao@torontomu.ca, vivian@torontomu.ca)}
\thanks{Guanghui Wang with the Department of Computer Science, Toronto Metropolitan University, Toronto, Canada (e-mail: wangcs@torontomu.ca).}}

\markboth{Journal of \LaTeX\ Class Files,~Vol.~18, No.~9, September~2020}%
{How to Use the IEEEtran \LaTeX \ Templates}

\maketitle

\begin{abstract}
Multi-modality magnetic resonance imaging (MRI) is essential for the diagnosis and treatment of brain tumors. However, missing modalities are commonly observed due to limitations in scan time, scan corruption, artifacts, motion, and contrast agent intolerance. Synthesis of missing MRI has been a means to address the limitations of modality insufficiency in clinical practice and research. However, there are still some challenges, such as poor generalization, inaccurate non-linear mapping, and slow processing speeds. To address the aforementioned issues, we propose a novel unified synthesis model, the \textbf{Frequency-guided and Coarse-to-fine Unified Diffusion Model (FgC2F-UDiff)}, designed for multiple inputs and outputs. Specifically, the Coarse-to-fine Unified Network (CUN) fully exploits the iterative denoising properties of diffusion models, from global to detail, by dividing the denoising process into two stages—coarse and fine—to enhance the fidelity of synthesized images. Secondly, the Frequency-guided Collaborative Strategy (FCS) harnesses appropriate frequency information as prior knowledge to guide the learning of a unified, highly non-linear mapping. Thirdly, the Specific-acceleration Hybrid Mechanism (SHM) integrates specific mechanisms to accelerate the diffusion model and enhance the feasibility of many-to-many synthesis. Extensive experimental evaluations have demonstrated that our proposed FgC2F-UDiff model achieves superior performance on two datasets, validated through a comprehensive assessment that includes both qualitative observations and quantitative metrics, such as PSNR SSIM, LPIPS, and FID. The source code is available at \url{https://github.com/xiaojiao929/FgC2F-UDiff}.
\end{abstract}


\section{Introduction}
\label{sec:introduction}

\begin{figure}[t]
\center
\includegraphics[width=0.5\textwidth]{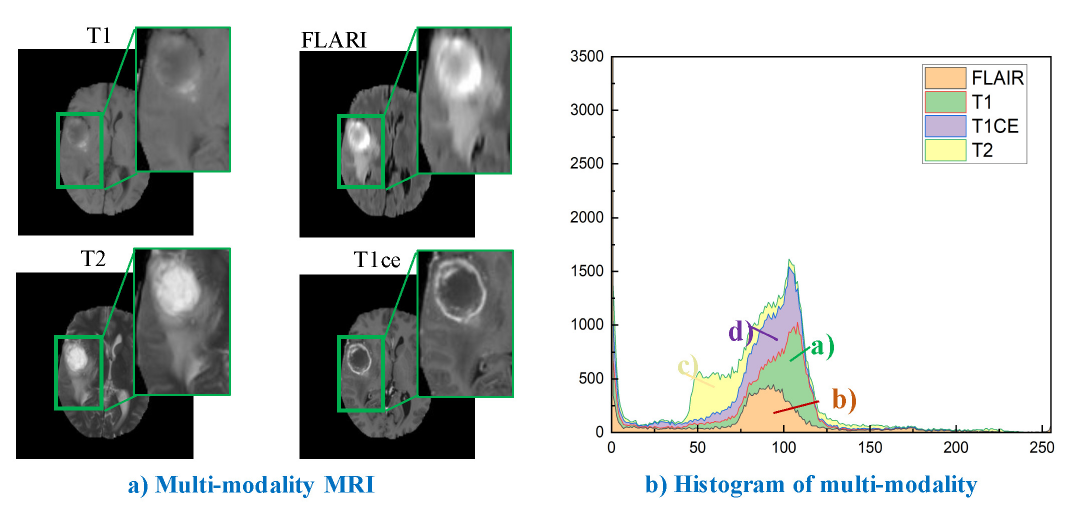}
\caption{Different image modalities provide different types of tissue contrast views and spatial resolution, which has variety in the histogram.} \label{fig1}
\end{figure}

\IEEEPARstart{M}{ulti-modality} magnetic resonance imaging (MRI), encompassing T1, T2, FLAIR, and T1 contrast-enhanced (T1ce) sequences, is indispensable for the diagnosis, monitoring, and treatment of brain tumors, as it provides complementary information about tissue views and spatial details \cite{menze2014multimodal,bagci2013joint,xiao2023edge}. As shown in Fig.\ref{fig1}, T1 images provide anatomical structure, FLAIR highlights the entire tumor region, T2 delineates a clear outline of the tumor edema area, and T1ce depicts a clear boundary of enhanced areas. However, the absence of modalities across different clinical centers is unavoidable \cite{kronberg2022optimal}. Furthermore, modalities are often missing due to limitations in scan time, scan corruption, artifacts, motion, and contrast agent intolerance \cite{chartsias2017multimodal,varsavsky2018}. These factors restrict the ubiquity of multi-modality, adversely affecting various downstream tasks such as segmentation, detection, and quantification. Consequently, the synthesis of missing multi-modality MRI has garnered increasing research interest as a means to overcome the limitations of modality insufficiency in clinical practice and research.

In recent years, numerous studies have provided evidence of the effectiveness of deep learning in synthesizing missing modality \cite{isola2017image,Zhao2020,huang2019coca,yu2019ea,yuan2020unified,cao2020auto,wang2024mix,yurt2021mustgan,xu2024disentangled}. Task-specific models, including one-to-one and many-to-one, focus on learning individual non-linear mappings from source to target imaging modalities, neglecting unique information present in shared features. These limitations can compromise synthesis performance and fail to meet the clinical needs for multi-modality applications. GAN-based methods utilize adversarial losses to better capture organizational structure and further enhance synthesis quality. Moreover, some unified method of GAN-based achieves many-to-many synthesis \cite{sharma2019missing,dalmaz2022resvit,zhang2023unified} to address the limitation of one modality synthesis. However, GAN-based methods suffer from mode collapse, non-convergence, instability, and high sensibility to hyperparameters \cite{bau2019seeing,dhariwal2021diffusion}.

Denoising diffusion probabilistic models (DDPM) may stably generate high-quality dependable images to improve MRI synthesis as a promising alternative to GAN \cite{ho2020denoising}. Diffusion models show superiority in a wide variety of areas, ranging from generative modeling tasks (e.g., image generation \cite{Zhao2024}, image super-resolution \cite{li2022srdiff,gao2023implicit,chao2023lfsrdiff}) to discriminative tasks (e.g., image segmentation \cite{rahman2023ambiguous,patel2024multi}, classification \cite{Zhao2021,vyas2024predicting}, and anomaly detection \cite{wolleb2022diffusion,Xiao2022}) while reducing risk of modality collapse. The work \cite{jiang2023cola} proposed a conditioned latent diffusion model for many-to-one synthesis, which preserves anatomical structure with accelerated sampling. However, only learning the mapping of a single task (i.e., many to one) is unable to meet clinical needs for multi-modality. And, the long time steps (t=1000) are unable to solve the slow sampling problem of DDPM \cite{croitoru2023diffusion}. Besides, significant challenges persist that limit the widespread application of existing methods: 1) Lack of an effective unified model to synthesize high-fidelity images from many to many. The existing methods suffer from poor generalization ability for atypical anatomical features, which limits the fidelity of synthesis, especially details. 2) Lack of effective strategy to guide the learning of unified highly non-linear mapping between multi-modality MRI. During the iterative denoising process, the randomness of noise alters the original distribution of the target image \cite{lyu2022conversion}. During the iterative denoising process, the randomness of the noise changes the original distribution of the target image. Therefore, the learned nonlinear mapping relationships are unable to accurately reflect the consistency of anatomical structures. 3) Lack of feasibility of many-to-many because of the slow sampling problem of DDPM. 

\textbf{Motivation}: To address the above-mentioned challenges, we propose a unified diffusion model, the Frequency-guided and Coarse-to-fine Unified Diffusion Model (FgC2F-UDiff) to synthesize the missing modality for multiple inputs and outputs. Our key hypothesis is to decompose the frequency domain into low-high-frequency as guidance information in coarse-to-fine stages based on the iterative denoising properties of diffusion models. Specifically, FgC2F-UDiff relies primarily on low-frequency with global anatomical features to guide coarse denoising and subsequently performs fine denoising progress guided by high-frequency with texture and detail. FgC2F-UDiff enables the effective synthesis of target modalities with high fidelity from any combination of modalities.

\textbf{Coarse-to-fine Unified Network (CUN)} fully utilizes iterative denoising properties of diffusion model (global-to-detail), novelty dividing the denoising process into two stages (i.e., coarse-to-fine) to improve the fidelity of denoised images. Moreover, CUN based on the diffusion model provides an effective unified program for cross-modality missing synthesis of multiple inputs and outputs.

\textbf{Frequency-guided Collaborative Strategy (FCS)} guides the learning of accurate non-linear mapping by enhancing the diversity of prior knowledge. Specifically, inspired by image signals, we incorporate low-frequency with global anatomical features in the early stages of coarse denoising and introduce high-frequency information with local fine-grained (i.e., texture and details) in the later stages of fine denoising. Therefore, frequency domain information of different granularity is decomposed and used to guide denoising collaboratively as prior knowledge, so the synthesized image has diverse features and similar edge and detail characteristics to the real image. At the same time, novel strategies were designed to dynamically search for appropriate frequency domain information to maximize the information of available modalities.

\textbf{Specific-acceleration Hybrid Mechanism (SHM)} is designed for specific tasks to accelerate the diffusion model and improve the fidelity of synthesized images. First, the curriculum learning (CL) mechanism is employed to simulate the easy-to-hard learning of missing modalities. Second, the network divided into two coarse-to-fine phases fits the iterative denoising properties of diffusion models, thus accelerating the synthesis process from the whole image to the details. Finally, dynamically selecting constraint conditions of frequency ensures the maximization of information from available modalities, guaranteeing the learning of non-linear mapping of image texture and fine-detail structures. Consequently, the trained model is adaptable to any number of original modalities and exhibits increased robustness in specific complex regions of images, enhancing the feasibility and synthesis performance of many-to-many FgC2F-UDiff. 

Our contributions include the following:

\begin{itemize}
\item{To the best of our knowledge, this is the first work to introduce a unified diffusion model guided by a frequency domain, which provides an effective cross-modality synthesis mechanism for multiple inputs and outputs.}
\item{We propose an innovative frequency-domain-guided coarse-to-fine network that effectively incorporates the iterative denoising characteristics of the diffusion model. This approach strategically shifts guidance across the appropriate frequency domains from coarse to fine, enhancing the fidelity of synthesized images.}
\item{We propose an efficient mechanism, SHM, which intelligently blends specific mechanisms to accelerate the diffusion model and improve the feasibility of many-to-many.}
\end{itemize}

\section{Related Work} \label{2.1}

Multi-domain synthesis of medical images provides a promising solution to address the limitations of modality insufficiency, which has attracted significant interest and gained popularity in recent years. Many research works and various technologies have been presented in the multi-domain synthesis of medical images. This section briefly reviews known synthesis methods by categorizing them into task-specific models (i.e., one-to-one and many-to-one) and unified models (i.e., many-to-many).

\subsection{Task-specific model for missing image synthesis}
    
\paragraph{\textbf{One-to-one}} Earlier one-to-one studies have proposed patch-based regression \cite{torrado2016fast,roy2016patch,roy2013magnetic}, sparse dictionary representation \cite{bowles2016pseudo, jog2017random}, and atlas \cite{jog2015mr,ye2013modality}. However, handcrafted features constrain the performance and development of these traditional methods. To improve the automatic extract feature, deep learning (DL) has been employed in cross-modality synthesis \cite{van2015cross,sevetlidis2016whole,li2014deep}. For instance, the work of \cite{van2015cross} developed a patch-based location-sensitive deep network (LSDN), which combines intensity and spatial information for synthesizing T2 MRI from T1 MRI and vice versa. The work of \cite{sevetlidis2016whole} proposed a deep encoder-decoder image synthesizer (DEDIS) for whole image synthesis. Despite yielding enhancements, CNN-based has the drawback of losing detailed structural information \cite{isola2017image}. GAN-based achieved great success with the development of deep learning techniques \cite{yuan2020unified,huang2019coca,yu2019ea,peng2021multi,yurt2022semi}. For instance, the work of \cite{huang2019coca} designed CoCa-GAN for synthesizing MRI data (i.e., T2, FLAIR, and T1ce) from T1, which utilizes adversarial learning and context-aware learning to learn common feature spaces.  The work of \cite{yuan2020unified} proposed a unified GAN, which learns the modality-invariant features by modality translation for segmentation tasks.

\paragraph{\textbf{Many-to-one}} Earlier many-to-one studies have proposed patch-based regression \cite{jog2014random,jog2017random}. Later studies also used DL-based, which also achieved great success \cite{mehta2018rs,chen2019robust,zhou2020hi}. For instance, the work of \cite{zhou2020hi} designed a Hybrid-fusion Network (Hi-Net) for multi-modal MR image synthesis, which employed a fusion network to learn the common latent representation of multi-modality data. The work of \cite{chen2019robust} proposed a multi-modality synthesis framework, which fused disentangled content code from each modality into a shared representation via gated feature fusion. Recently, GAN-based methods were demonstrated to outperform other DL-based methods in many-to-one tasks \cite{yu2019ea,lee2019collagan,lee2020assessing,yurt2021mustgan,xu2021domain,zhan2021multi,yurt2022progressively}. For instance, the work of \cite{yu2019ea} presents Ea-GANs to generate T2 and FLAIR images from T1. The Ea-GAN captured the edges of key texture information, and two GAN variants are proposed to integrate the edge information through different learning strategies. Lee et. al\cite{lee2019collagan} proposed a CollaGAN framework for missing image data imputation,  which converts the image imputation problem to multi-domain images-to-image translation tasks. The work of \cite{cao2020auto}generated any missing modality in a single unified Auto-GAN model, which performs self-supervised learning to learn multi-facet information, further guaranteeing its generalizability. 

However, when an insufficient number of modalities are available, especially when many modalities (e.g., three modalities) do not exist, applying the above strategies can not necessarily ensure that the lost data is recovered since there are not enough features to reconstruct the missing data \cite{ding2021rfnet}.

\subsection{Unified model for missing image synthesis}

Unified synthesis methods take multiple inputs and generate multiple inputs and outputs, which is a relatively new study in synthesis tasks. Several studies have attempted to propose a unified model on many-to-many synthesis that can exploit all available data. For instance, the work of \cite{sharma2019missing} proposed a multi-modality generative adversarial network (MM-GAN), which was one of the first to propose a multi-input and multi-output architecture that generalizes to any combination of available and missing modalities. The work of  \cite{dalmaz2022resvit} proposed an adversarial model with a residual vision transformers (ResViT) generator to translate between multi-modal imaging data. The work of \cite{zhang2023unified} exploits the commonality information of available modalities for unified multi-modal image synthesis based on GAN. However, the above methods are all based on GAN \cite{mirza2014conditional}, which has some common issues while training GAN, such as mode collapse, non-convergence, instability, and high sensibility to hyperparameters, thus limiting the fidelity and diversity of synthesized images \cite{bau2019seeing,dhariwal2021diffusion}. The work of \cite{meng2024multi} proposed unified Multi-modal Modality-masked Diffusion Network (M2DN), tackling multi-modal synthesis from the perspective of “progressive whole-modality inpainting”, instead of “cross-modal translation”. However, it only takes the available modes as conditions and does not take into account the denoising iterative properties of the diffusion model.

\section{Methodology}

\begin{figure*}[t]
\center
\includegraphics[width=0.9\textwidth]{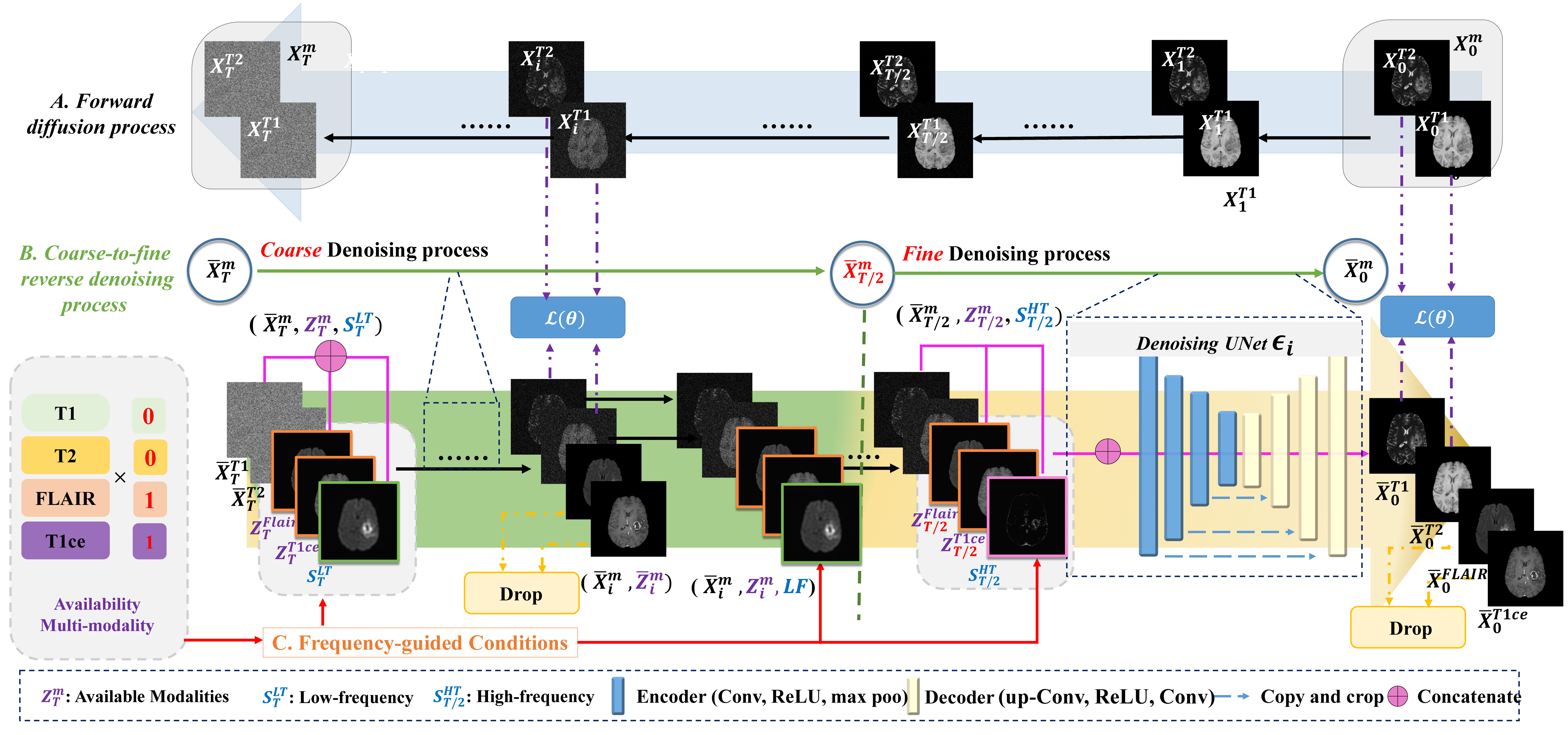}
\caption{Overview of the proposed FgC2F-UDiff, which cross-modality synthesizes missing modalities from multiple inputs and outputs. It includes forward diffusion progress and coarse-to-fine reverse denoising progress. The FgC2F-UDiff decomposes the frequency domain into low-high-frequency as guidance information in coarse-to-fine stages based on the iterative denoising properties of diffusion models.} \label{figure2}
\end{figure*}

As shown in Fig.\ref{figure2}, the proposed FgC2F-UDiff integrates any available number of source image modalities (i.e., T1, T2, FLAIR, and T1ce) for coarse-to-fine synthesizing missing target modalities $ X_{0}^{m} \in \mathbb{R}^{H\times W}$, where $H$ and $W$ represent the height and width, respectively. The FgC2F-UDiff works via a forward diffusion process in Section \ref{forward} and a coarse-to-fine reverse denoising process. Specifically, the coarse-to-fine unified network (CUN) divides the denoising process into two stages of coarse-to-fine for improving the fidelity of synthesizing image in Section \ref{CUN}. Among them, the frequency-guided collaborative strategy (FCS)  decomposes the frequency information to guide the learning non-linear mapping of many-to-many. It utilizes low-frequency and high-frequency to enhance the realism of the synthesized image structures in Section \ref{FCS}. The entire network benefits from a specific-acceleration hybrid mechanism (SHM) to accelerate the time steps to improve the availability of many-to-many FgC2F-UDiff in Section \ref{SHM}.

\subsection{Forward diffusion process} \label{forward}
The forward diffusion process of FgC2F-UDiff is defined as a Markov chain as DDPM \cite{ho2020denoising}, which maps between source samples and pure noise samples. Formally, given the multi-modality of $S_{0}^{m}$, $m \in \{T1, T2, FLAIR, T1ce\}$, among missing-images are  $X_{0}^{m} \in {S_{0}^{m}}$, $X_{0}^{m} \sim  \mathbb{Q}  (X_{0}^{m}$). The FgC2F-UDiff gradually adds Gaussian noise to  $X_{0}^{m}$ and obtains a pure noise sample $X_{T}^{m}$ with time steps  $T \in \{1,2,...t, t+1, ..., T-1, T\}$. At the time step $t$, the noisy $X_{t}^{m}$ can be formulated as:
\begin{equation}
 X_{t}^{m} =  \sqrt{1- \beta _{t}} X_{t-1}^{m}+  \beta _{t} \epsilon_{t-1}   \\\label{eq1}
\end{equation}
where $\beta_{t} \sim (0, I)$ is the variance of the Gaussian noise added at time step $T$, $ \epsilon \sim \mathcal{N} (0,1)$ is Gaussian distribution noise. Thus, the forward diffusion process can be formulated as:

\begin{equation}
 \mathbb{Q} ( X_{t}^{m} | X_{t-1}^{m} ) := \mathcal{N} ( X_{t}^{m}; \sqrt{1- \beta _{t}} X_{t-1}^{m},  \beta _{t} I)   \\ \label{eq2}
\end{equation}
where $I$ denotes the standard normal distribution. Using the notation $a_{t} := 1- \beta _{t}$ and  $ \bar{a_{t}}= \prod_{s=1}^{t}a_{s}$, the forward process admits sampling $X_{t}^{m}$ at an arbitrary timestep $t$ can be deduced by Eq.\ref{eq1} and  Eq.\ref{eq2}: 

\begin{equation}
 \mathbb{Q} ( X_{t}^{m} | X_{0}^{m} ) = \mathcal{N}  ( X_{t}^{m};  {\sqrt{\bar{a_{t}}}} X_{0}^{m} , (1-\bar{a_{t}}) I)  \\
\end{equation}

\subsection{Coarse-to-fine Unified Network (CUN)} \label{CUN}

To improve the performance of the many-to-many synthesis, FgC2F-UDiff designed a coarse-to-fine collaborative reverse diffusion process as shown in Fig.\ref{figure2}. Specifically, in the light of the iterative denoising properties of diffusion models (global-to-detail) in the diffusion model, our reverse diffusion process is divided into two phases: a coarse denoising process from $T$ to $\frac{T}{2}-1$, followed by a fine denoising process from $\frac{T}{2}$ to 0. Simultaneously, FgC2F-UDiff leverages the characteristics of image signals (high- and low-frequency), by decomposing the frequency domain to guide the synthesis in a staged manner. Additionally, FgC2F-UDiff fully leverages all available modalities as conditions to enhance and expedite the denoising process. As a result, our FgC2F-UDiff possesses the capability to cross-modality synthesis mechanism for multiple inputs and outputs. 

\textbf{Reverse denoising diffusion process.} Since the reverse of the forward process is intractable, DDPM learns parameterized Gaussian transitions. Given $X_{t}^{m}$ and corresponding conditional of all conditions $C_{t}^{m}$, in each time step of the reverse process, the denoising operation is performed on the noisy multi-channel image  ($X_{t}^{m}$, $C_{t}^{m}$) to obtain the previous image $X_{t-1}^{m}$. The probability distribution of $X_{t-1}^{m}$ under the condition $X_{t}^{m}$ can be formulated as: 
\begin{equation}
    \begin{split}
   P (X_{t-1}^{m} \! \! \mid X_{t}^{m} ,C_{t}^{m}) := \mathcal{N}  ( X_{t-1}^{m};  \mu _{\theta}(X_{t}^{m}, t,C_{t}^{m}),  \\
   \sigma _{\theta} (X_{t}^{m}, t,C_{t}^{m}) I )
    \end{split}
\end{equation}
where $\sigma _{\theta}$ is the variance of conditional distribution $P ( X_{t-1} ^ {m} \mid X_{t}^{m} , C_{t}^{m})$, which can be formulated as:

\begin{equation}
\sigma _{\theta}= \frac{1-\bar{a}_{t-1}}{1-\bar{a}_{t}}\beta _{t} \\
\end{equation}
\vspace{-0.5cm}
where $\beta _{t} = 1- a_{t}$. 
The generative process is expressed as: 

\begin{equation}
    \begin{split}
    X_{t-1}^{m} = \frac{1}{\sqrt{\bar{a_{t}}}}( X_{t}^{m} - \frac{\beta _{t}}{\sqrt{1-\bar{a_{t}}}} \epsilon_{\theta}(X_{t}^{m},t, C_{t}^{m}) + \\
\sigma _{\theta}(X_{t}^{m},t, C_{t}^{m}) Z, Z \sim \mathcal{N} (0,1)
    \end{split}
\end{equation}
where $\epsilon_{\theta}$ represents noise approximation.

\begin{figure}[t]
\center
\includegraphics[width=0.5 \textwidth]{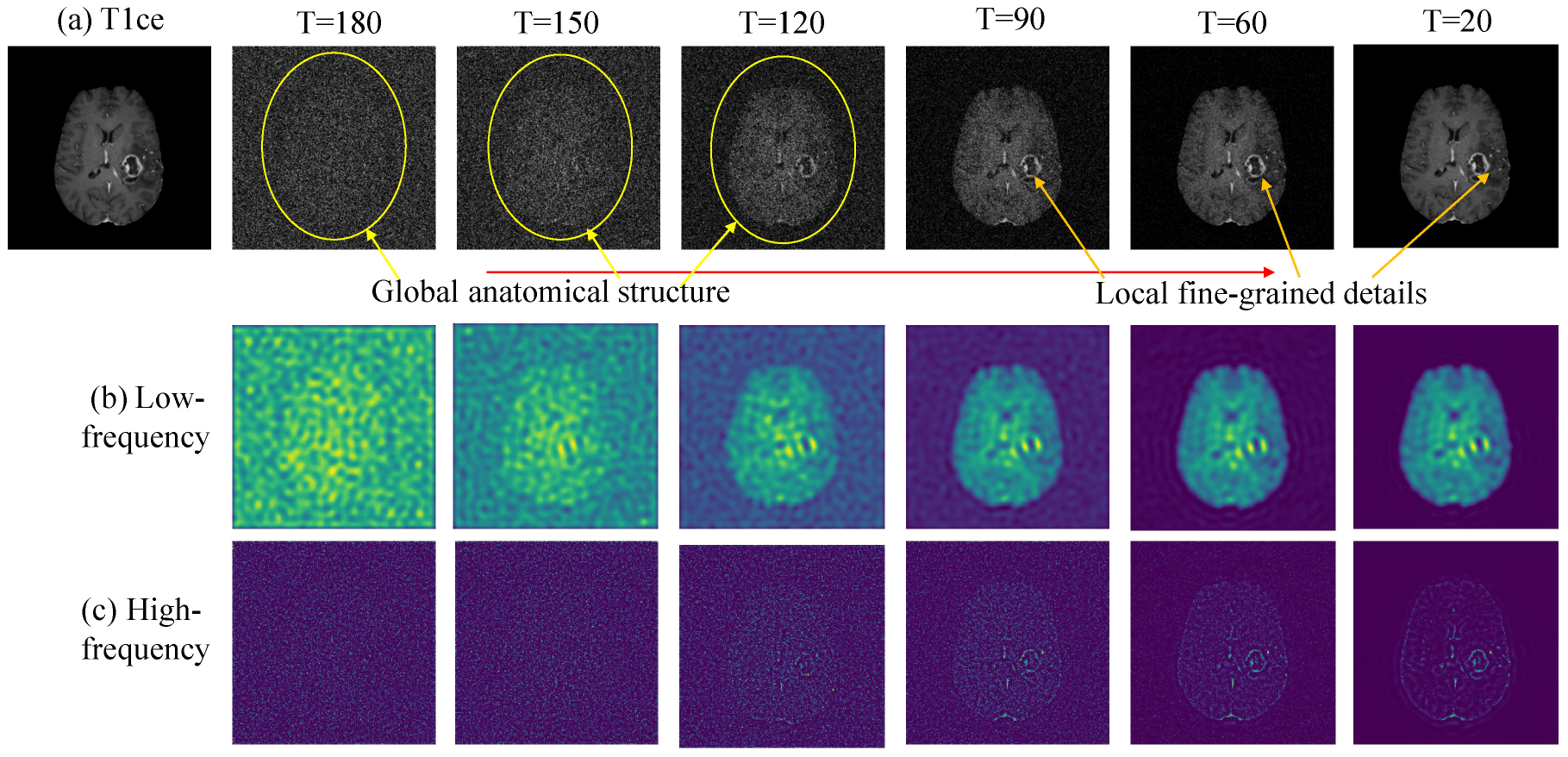}
\caption{Visualize analyzing and visualizing the denoising synthesis images, which shows the significant properties of iterative denoising. Specifically, (a) shows denoised synthesis images corresponding to different time steps $T$. (b) shows the low frequency. (c) shows the high frequency.} \label{diffusion}
\end{figure}

 Denoising models based on the UNet \cite{ronneberger2015u}, which is widely used in a diverse range of segmentation and synthesis tasks due to the U-shaped symmetrical structure and skip connection between the encoder and decoder. The architecture is shown in Fig.\ref{figure2}. FgC2F-UDiff is trained to predict a denoised variant of their input $X_{T}^{m}$, where $X_{t}^{m}$ is a noisy version of the input $X_{0}^{m}$. Specifically, the network is designed with a connection input ($ \mathbb{C} _{T} \in \mathbb{R}^{(H\times W)\times N_{c}} $) with 5-channels, where channel $N_{c}$ = 0,1,2,3 and 4 corresponds to $T1, T2, FLAIR, T1ce$ and frequency-guided image, respectively. The input image $ \mathbb{C} _{T}$ connects all missing-images $X_{T}^{m}$, and conditional images $C_{T}^{m}$ (${Z_{T}^{m}}$, $S^{LF}$ or $S^{HF}$). In the frequency-guided conditions module (as shown in Fig.\ref{figure2}.C), all modalities $S_{0}^{m}$ have been sorted according to $T1$, $T2$, $FLAIR$, $T1ce$ to obtain the corresponding 4-digit numbers, the numeral ``1'' signifies the presence of a particular module, while ``0'' indicates its absence. After $S_{0}^{m}$ multiplied by the corresponding digit number, the channels corresponding to each missing modality are inputted with noisy images of $X_{T}^{m}$, the channels corresponding to each condition modality are inputted with source images as ${Z_{T}^{m}}$, and the channel corresponding to conditional frequency is inputted with $S^{LF}$ during coarse denoising process, while $S^{HF}$ during fine denoising process. For instance, if sequences $T1$ and $T2$ are missing, channels $N_{c}$=0 and $N_{c}$=1 are fed with noisy $X_{T}^{T1}$ and $X_{T}^{T2}$, respectively. $N_{c}$=2 and $N_{c}$=3 are fed with original images ${Z_{T}^{FLAIR}}$ and ${Z_{T}^{T1ce}}$, respectively. And, the last $N_{c}$=4 is fed with corresponding frequency image $S^{LF}$. After one time step, UNet outputs 4 channels corresponding to 4 modalities. FgC2F-UDiff calculates the difference between the output $\bar{X}_{T-1}^{m}$ and ${X}_{T-1}^{m}$ to train the network, while copying the output $\bar{X}_{T-1}^{m}$ to enter the next iteration. To ensure effective guidance of available images, the synthesized $\bar{Z}_{T-1}^{m}$ is dropped. The encoder is a traditional stack of $3\times3$ convolution and $2\times2$ max pooling layers. And, the symmetric decoder is a traditional stack of $2\times2$ up-conv, copy, and crop, and $3\times3$ convolution. To modify network performance, each convolutional layer follows a BN layer and a ReLu layer.

  The FgC2F-UDiff is trained to synthesize the target modality by predicting the involved noise $\epsilon_{\theta }$ under the guidance of the $C_{t}^{m}$, which is formulated below:
\begin{equation}
     L _{FgC2F-UDiff} = E_{x_{t}, \epsilon \sim N(0,I),t}  \left\| \epsilon - \epsilon _{\theta} (X_{t}^{m},t,C_{t}^{m})\right\|_{2}^{2} \\
\end{equation}

\emph{\textbf{Discussion:} What is the iterative denoising properties of diffusion models (global-to-detail) in the FgC2F-UDiff.}

After analyzing and visualizing the denoising synthesis images, as shown in Fig.\ref{diffusion}, we can observe significant properties with iterative denoising. Specifically, (a) shows denoised synthesis images corresponding to different time steps $T$. In the early stage of denoising, it initially formed the global anatomical structure of the brain (as shown in the yellow circle). In the later stage, we are gradually synthesizing local fine-grained details such as the edges and texture of the tumor (as indicated by the orange arrow). These discrepancies gradually diminish from the global to the local details as the iteration of the denoising process $T$. (b) and (c) display the low-frequency and high-frequency images corresponding to the denoising image. In low-frequency images during the initial denoising phase ($T$ to $\frac{T}{2}-1$), the images exhibit clear and diverse global features, closely related to the anatomical structure of the brain. In high-frequency information images during the initial denoising phase ($\frac{T}{2}$ to 0), fine-grained features become increasingly evident and sharp, particularly around the edges of the brain and tumor. There is a higher demand for detailed features in later synthesis. Therefore, the proposed CUN fully considers the iterative denoising properties of diffusion models and the character of the frequency domain, divides denoising into two phases, and adds corresponding low-frequency and high-frequency information to guide the denoising process.

\subsection{Frequency-guided Collaborative Strategy (FCS)} \label{FCS}

To learn the high non-linear mapping of many-to-many, especially anatomical structures, we designed an FCS strategy guided by frequency. Specifically, Fig.\ref{diffusion} has verified the iterative denoising properties of diffusion models from global to local. So, FgC2F-UDiff incorporates low-frequency with global anatomical features in the early stages of coarse denoising from $T$ to $\frac{T}{2}-1$. And introducing high-frequency information with fine-grained (i.e., texture and details) as prior knowledge in the later stages of fine denoising from $\frac{T}{2}$ to {0}. The coarse and fine denoising stages work collaboratively in two phases to ensure the learning of the unified distribution of data and the production of high-quality synthesized images.

To find the most suitable frequency domain information as prior guidance knowledge, we have designed two strategies to dynamically search frequency information. Our FCS dynamically selects frequency information tailored to the available modalities at each stage of the different subject, employing a left-to-right scan for coarse structural low-frequency guidance and a right-to-left scan for fine detail high-frequency enhancement. This approach ensures optimal guidance for the diffusion process, adapting to the unique needs of each subject. Specifically, all modality $S_{0}^{m}$ have been sorted according to $T1$, $T2$, $FLAIR$, $T1ce$ to obtain the corresponding 4-digit numbers, the numeral ``1'' signifies the presence of a particular module, while ``0'' indicates its absence. As shown in Fig.\ref{figure2}.{\it B}, the missing modalities, when multiplied by zero, do not contribute any information to the denoising process, effectively excluding them from the calculation. Conversely, the available modalities are multiplied by one, preserving their original image data intact for further processing. Then, the ''left to right'' strategy extracts low-frequency by searching the corresponding modality of the first available image from left to right and defined as $S^{LF}$. And, the ''right to left'' strategy to extract high-frequency by searching the corresponding modality of the first available image from right to left and defined as $S^{HF}$.

To filter the images into their respective frequency domains, we apply Gaussian low pass filters (GLPF) \cite{heideman1984gauss} and Gaussian high pass filters (GHPF) \cite{schowengerdt1980reconstruction} to process $S^{LF}$ and $S^{HF}$, respectively. Specifically, we set that Gaussian kernel as: 

\begin{equation}
      \kappa_{\sigma}\left [ i,j \right ] = \frac{1}{2\pi \sigma^{2}} e^{-\frac{1}{2}(\frac{i^{2}+j^{2}}{\sigma ^{2}})} \\ \label{k}
\end{equation}
where $[i, j ]$ is the original point, and $\sigma$ is used to measure the width of the Gaussian curve. After the GLPF filter on $S^{LF}$, we obtain the low frequency image ($LF$):

\begin{equation}
      LF \left [ i,j \right ] = \sum_{m}\sum_{n} \kappa \left [ m,n \right ]\cdot S^{LF}\left [ i+m,j+n \right ] \\
\end{equation}
where $m, n$ represents the index of GLPF. Then, the high frequency image ($HF$) of $S^{HF}$ is expressed as: 

\begin{equation}
      HF  \left [ i,j \right ] =  1-   \sum_{m}\sum_{n} \kappa \left [ m,n \right ]\cdot S^{HF}\left [ i+m,j+n \right ]  \\
\end{equation}

\begin{figure}[t]
\center
\includegraphics[width=0.5 \textwidth]{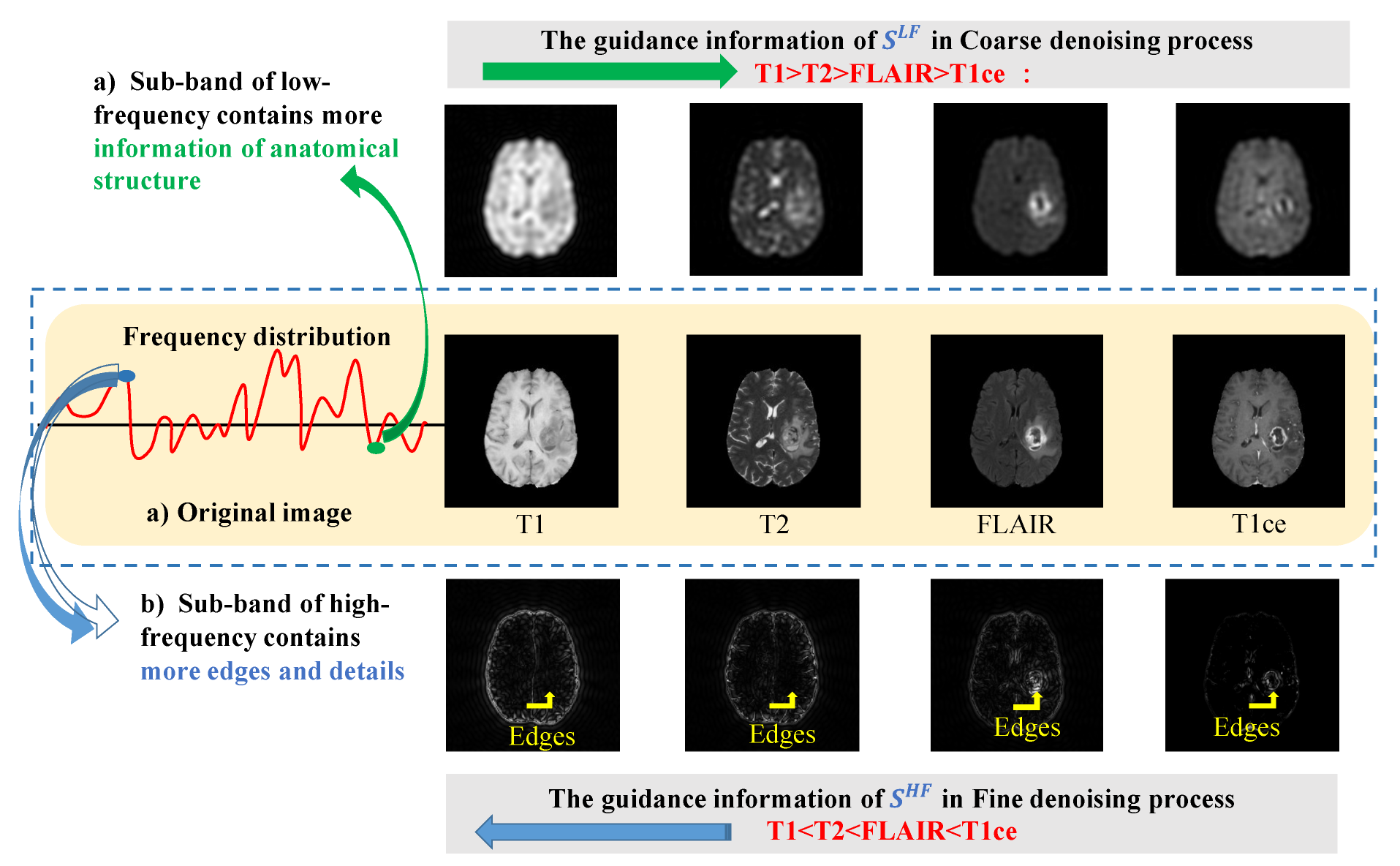}
\caption{The image can usually be decomposed into a high-frequency sub-band with edges and details and a low-frequency sub-band with anatomical structure.} \label{frequency}
\end{figure}

\emph{\textbf{Discussion:} Why design different strategies to dynamically search frequency domain information?}

As shown in Fig.\ref{frequency}, inspired by the character of image signals, the image can usually be decomposed into high-frequency sub-band and low-frequency sub-band. The high-frequency sub-bands contain more details and edge information, whereas the low-frequency sub-band contains the contour and structure information of images. And, the different modalities generated through scanning parameters usually provide different information \cite{bagci2013joint}. Such as T1 brain images delineated low-frequency with anatomical structure, while T1ce provides a high-frequency with a clear boundary between the enhanced areas. So, sorting the four modalities according to the amount of low-frequency information they contain can be obtained: $T1  > T2  > FLAIR  > T1ce$, whereas sorting the four modalities according to the amount of high-frequency information they contain can be obtained: $T1  < T2  < FLAIR < T1ce$. Therefore, we novelty designed two different selection methods according to the character of modality, that is, from ''left to right'' to find the best low-frequency information and from ''right to left'' to find the best high-frequency information. Through such a dynamic search strategy, the best guidance can be provided for the diffusion process.

\subsection{Specific-acceleration Hybrid Mechanism (SHM)} \label{SHM}

Our SHM is designed to accelerate the sampling speed of the diffusion model to improve the feasibility of many-to-many FgC2F-UDiff. Specifically, first, the curriculum learning (CL) mechanism is employed to simulate the easy-to-hard learning of missing modalities. Second, the coarse-to-fine denoising process guided by frequency fits the iterative denoising properties of diffusion models to accelerate the synthesis process from the whole image to the details. Third, dynamically selecting high-frequency and low-frequency constraint conditions ensures the maximization of information from available modalities. 

\textbf{Curriculum learning (CL) mechanism.} To ensure the adaptability of our FgC2F-UDiff for the synthesis of missing data across diverse inputs, we employ curriculum learning (CL) \cite{bengio2009curriculum} as a rationality-enhancing training strategy. Due to the varying degrees of data loss and difficulty in obtaining complete modalities, it is imperative to devise a training strategy that effectively leverages available data and expedites model convergence, ultimately yielding higher-quality synthesized results. CL shares some similarities with boosting algorithms, where the focus is gradually shifted towards more challenging examples. However, unlike a uniform distribution of training data, CL begins by emphasizing easier examples and progressively introduces more complex instances as the training process unfolds. In the context of CL-based training within our FgC2F-UDiff framework, we categorize the missing sequences into different difficulty levels. Specifically, we designate the task of synthesizing one missing sequence as ``easy'', followed by tackling the challenge of synthesizing two missing sequences categorized as ``moderate'', and finally, addressing the most demanding scenario of synthesizing all three missing sequences, denoted as ``hard''. This tiered approach to CL ensures that the model is systematically exposed to increasingly complex situations, enabling it to learn and adapt effectively across a range of input conditions.

 \section{Dataset and evaluation metrics}
 
 \subsection{Dataset} FgC2F-UDiff framework evaluated the performance on BraTS 2021 and IXI brain image datasets.
  \paragraph{ Brain Tumor Segmentation Challenge 2021 (BraTS 2021)} The BraTS 2021 \cite{baid2021rsna,menze2014multimodal,bakas2017advancing} contains 1,251 cases, consisting of four different MRI sequences per case (i.e., T1, T2, FLAIR, and T1ce), acquired with different protocols and various scanners from multiple institutions. Standardized pre-processing has been applied to all the sequences. Specifically, the dimension of each data is resampled to 240 $\times$ 240 $\times$ 150, and the intensity is normalized to the range $\left[-1,1 \right]$. More details about the preprocessing information can be found in the original publication \cite{menze2014multimodal}.
 \paragraph{Information Extraction from Images (IXI)} The IXI dataset \cite{IXI} contains nearly 600 MRIs from normal and healthy subjects, consisting of three different MRI sequences (i.e., T1, T2, and PD-weighted). The images were acquired with the following parameters (T1 image: TE = 4.603 ms, TR = 9.813 ms, spatial resolution = 0.94 $\times$ 0.94 $\times$ 1.2 $mm^{3}$, matrix size = 256 $\times$ 256 $\times$ 150. T2 image: TE = 100 ms, TR = 8178.34 ms, spatial resolution = 0.94 $\times$ 0.94 $\times$ 1.2 $mm^{3}$, matrix size = 256 $\times$ 256 $\times$ 150. PD-weighted image: TE = 8 ms, TR = 8178.34 ms, spatial resolution = 0.94 $\times$ 0.94 $\times$ 1.2 $mm^{3}$, matrix size = 256 $\times$ 256 $\times$ 150). Note that the multi-contrast images in this dataset were unregistered. Therefore, T2 and PD-weighted images were spatially registered onto T1-weighted images before modeling by rigid transformation. Registration was performed via an affine transformation in FSL \cite{jenkinson2001global} based on mutual information.

\subsection{Implementation details} We employed five-fold cross-validation to train and test the grading. For each cross-validation split, the dataset was divided into a training/validation/testing as 7:1:2. Our network was implemented on Ubuntu 20.04 platform, using Python v3.6 and PyTorch v0.4.0, and was run on 2 NVIDIA GTX 3090Ti GPUs with 24 GB memory. FgC2F-UDiff are optimized using Adam optimizer \cite{kingma2014adam} with a learning rate of 0.0001. Following the noise schedules of DDPM \cite{ho2020denoising} and set the value of time steps $T$ to 200. Following the work of \cite{schowengerdt1980reconstruction}, the kernel size $\kappa$ in Eq.(\ref{k}) was set to 21. 
 
\subsection{Evaluation index} The performance of FgC2F-UDiff is evaluated by four standard measures, including peak signal-to-noise ratio (PSNR), structural similarity index (SSIM) \cite{wang2004image}, learned perceptual image patch similarity(LPIPS), and fréchet inception distance (FID), which reflect the quality and variety of synthetic images. The significance of performance differences was evaluated with signed-rank tests (p $<$ 0.05). The evaluation criteria are defined as: 
\begin{equation}
  PSNR=10\cdot{log}_{10}(\frac{MAX_I^{2}}{\sqrt{MSE}})
\end{equation}
\begin{equation}
  SSIM=\frac{(2Avg_xAvg_y+C1)(2 \delta_{xy}+C2)}{(Avg_x^2+Avg_y^2+C1)(  \delta_x^2 \delta_y^2+C2)}
\end{equation}
where $MAX_I$ represents the max value among the pixels in brain MRI with a size of $m\times{n}$. $x$ and $y$ are two images to be compared. $Avg_x$ and $Avg_y$ are the average pixel value of $x$ and $y$, respectively. $ \delta_x^2$ and $ \delta_y^2$ are the variance of $x$ and $y$, respectively. And $ \delta_{xy}$ is the covariance of $x$ and $y$. $C1=(k_{1}L)^{2}$, and $C2=(k_{2}L)^{2}$ are two constants, avoiding division by zero. $L$ is the range of pixel values, $k_{1}=0.01$ and $k_{2}=0.03$ are the default.
\begin{equation}
  \text{LPIPS}(x, y) = \sum_{l} w_l \cdot \frac{1}{H_l W_l} \sum_{h,w} \| \phi_l(x)_{h,w} - \phi_l(y)_{h,w} \|_2^2
\end{equation}
where $\phi_l(x)$ and $\phi_l(y)$ are the feature activations at layer $l$ of the network for images $x$ and $y$, respectively. $H_l$ and $W_l$ are the height and width of the feature maps at layer $l$. And, $w_l$ are the learned weights for the features at layer $l$, optimized to align with human perceptual difference. $\|.\|_{2}$ denotes the Euclidean distance. 

\begin{equation}
    \text{FID}(p, q) = \| \mu_p - \mu_q \|^2 + \text{Tr}(\Sigma_p + \Sigma_q - 2(\Sigma_p \Sigma_q)^{1/2})
\end{equation}
where $p$ and $q$ represent the distributions of features extracted from the real and generated images, respectively. And, $\mu_p$, $\mu_q$ are the mean vectors of the features from distributions $p$ and $q$. $\Sigma_p$, $\Sigma_q$ are the covariance matrices of the features from distributions $p$ and $q$. \text{Tr} denotes the trace of a matrix, which is the sum of the elements on the main diagonal. $\Sigma_p \Sigma_q^{1/2}$ represents the square root of the product of the covariance matrices, used to calculate the similarity between the two distributions. In this study, the FID was computed using a singular comprehensive evaluation of the model-generated images against the reference dataset.

\subsection{Comparison settings} To demonstrate the superiority of our proposed framework, FgC2F-UDiff is compared with other methods on two datasets. The baseline methods include the task-specific models (pix2pix \cite{isola2017image}, pGAN \cite{dar2019image}, LDM \cite{rombach2022high}, and CoLa-Diff \cite{jiang2023cola}) and unified models (MM-GAN \cite{sharma2019missing}, ResVit \cite{dalmaz2022resvit}, and Uni-GAN \cite{zhang2023unified}). The hyperparameters of each competing method were optimized via identical cross-validation procedures.

\section{Experiments}

The experiment results show that FgC2F-UDiff achieves high performance in both task-specific and unified synthesis regarding PSNR, SSIM, LPIPS, and FID. A set of experiments were performed to evaluate the performance of FgC2F-UDiff, including (1) synthesis results of the proposed FgC2F-UDiff in Section \ref{syn}; (2) performance comparison of task-specific synthesis with state-of-the-art (SOTA) methods in Section \ref{task}; (3) performance comparison of unified synthesis models with state-of-the-art (SOTA) methods in Section \ref{uni}; (4) ablation studies of FgC2F-UDiff in Section \ref{ablation}; and (5) analysis of SHM in Section \ref{SHM1} 

\begin{figure*}[t]
\center
\includegraphics[width=0.8\textwidth]{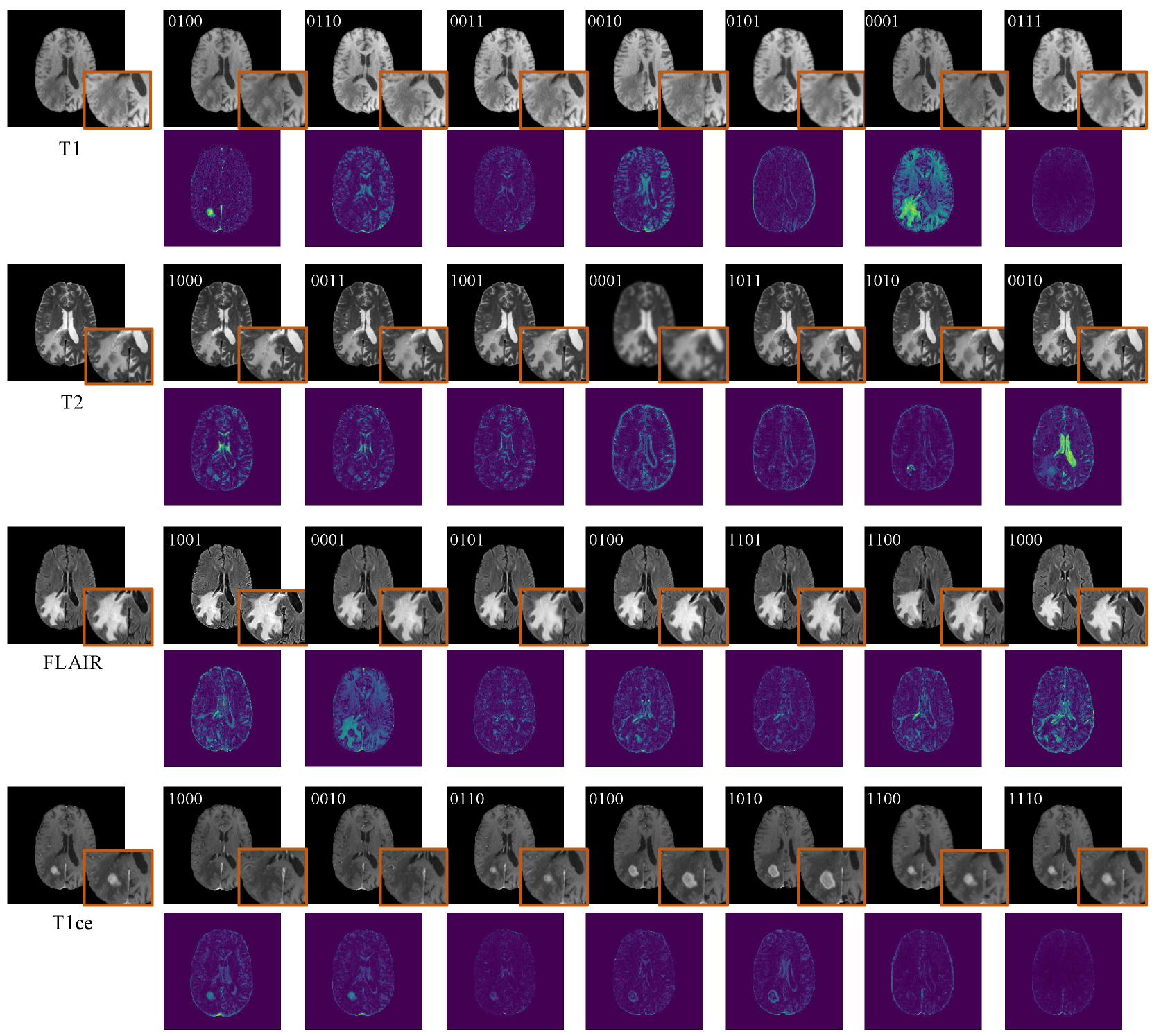}
\caption{Illustrative instances of synthetic images generated by our FgC2F-UDiff on the BraTS Dataset. Each row shows composite diagrams portraying distinct modes and error maps juxtaposed with the corresponding ground truth. The enlarged orange squares represent selected regions with notable disparities, providing enhanced insights into texture, edge enhancement, and shape characteristics.} \label{Our_Brats}
\end{figure*}

\subsection{Synthesis results of the proposed method} \label{syn}

The visual qualitative results of FgC2F-UDiff are shown in Fig.\ref{Our_Brats}. The four-bit digits in the figures indicate the availability conditions of T1, T2, FLAIR, and T1ce modalities. The digit ``1'' signifies the availability of a particular modality, while ``0'' indicates its absence. Specifically, for T1 sequences, the synthetic results derived from multiple sequences (i.e., 0111) yield the most faithful quality with minimal error compared to the ground truth. This contrasted starkly with single-sequence inputs (i.e., 0100, 0010 and 0001). These multi-sequence inputs yielded synthetic results with diminished noise levels and sharply defined boundaries between white and gray matter regions. It is noteworthy that a clear anatomical structure, with less information about the tumor enhancement area, characterizes the T1 image. Therefore, it is easy to lose tumor enhancement areas without the guidance of complementary information from other modalities, such as 0110. Because T1ce modality provides a clear boundary between the regions enhanced around the tumor. Visual results eloquently establish the indispensability of integrating complementary information from diverse modalities to achieve precision in synthesizing tumor regions with accurate shapes and realistic textures. The synthetic sequence quality exhibits significant improvement with the increased number of available input sequences. This improvement is attributed to the inherent diversity and complementary nature of information encapsulated within different modalities. Meanwhile, the differential outcomes obtained across these scenarios vividly illustrate different modalities' varying contributions to the target sequence synthesis process. Furthermore, these results validate our model's ability to generalize effectively when confronted with varying quantities of available modalities.

The quantitative results of FgC2F-UDiff are summarized in Table.\ref{Brats_syn} and Table.\ref{IXI_syn}. The values in the tables represent the average results of using the input modality to synthesize all other target modalities. Specifically, On the BraTS dataset (as shown in Table.\ref{Brats_syn}), the first column (number 1) indicates the input modality (e.g., T1). The PSNR of 26.13 $dB$ and SSIM of 0.887 values represent the average results of using the input modality to synthesize all other target modalities (e.g., T2, FLAIR, and T1ce). Compared with it, the combination of T1 and T2 (number 6) further improves the synthesis quality by 1.46 $dB$ and 0.032 in terms of PSNR and SSIM, respectively. In addition, the combination of T2 and FLAIR information (number 11) dramatically improves the synthesis quality by 2.19 $dB$ and 0.055 for PSNR and SSIM, respectively. On IXI dataset (as shown in Table.\ref{IXI_syn}), FgC2F-UDiff achieves 34.00 $dB$ in PSNR and 0.967 in SSIM using single T1 (number 1). Compared with it, the combination of T1 and T2 (number 4) improves the synthesis quality by 0.72 $dB$ and 0.005 for PSNR and SSIM, respectively. The quantitative results indicate that the integration of the maximum number of accessible modalities attains optimal performance. This consistency between quantitative and qualitative findings underscores the significance of leveraging multiple modalities for enhanced synthesis outcomes. Besides, the synthesis results obtained in our experiments exhibit substantial variations across different datasets. These disparities are primarily attributed to two key factors. First, the inherent divergence in imaging principles gives rise to fundamental differences in the information encapsulated within the images themselves. Second, the complementary information embedded within different combinations of modalities leads to divergent guidance for the synthesis process. Consequently, the differential contributions of these modalities result in pronounced disparities in the quality and fidelity of the synthesized outputs.

\begin{table}[h]
  \centering
   \setlength \tabcolsep{4pt}
\caption{Quantitative results of our method on the BraTS dataset. PSNR and SSIM are reported values are mean $\pm$ std.}\label{Brats_syn}
\begin{tabular}{llllllll}
\hline
&\multicolumn{4}{c}{Available modalities} & \multicolumn{2}{c}{Results} \\
Number & T1 &FLAIR & T2 & T1ce &  PSNR($dB$)   &    & SSIM   \\\hline
1&$\checkmark$ &   &   &  & 26.13$\pm$1.32  && 0.887$\pm$0.014  \\\hline
2&& $\checkmark$  &   &   & 25.96$\pm$1.51  && 0.881$\pm$0.015  \\\hline
3&&   &  $\checkmark$ &   & 26.79$\pm$1.44  && 0.896$\pm$0.012 \\\hline
4&&&&$\checkmark$ & 27.12$\pm$1.09  && 0.903$\pm$0.017  \\\hline
5&$\checkmark$& $\checkmark$  &   &   & 27.38$\pm$1.32 && 0.917$\pm$0.015  \\\hline
6&$\checkmark$&   & $\checkmark$  &   & 27.59$\pm$1.24 && 0.919$\pm$0.009  \\\hline
7&$\checkmark$&   &   & $\checkmark$  & 27.93$\pm$1.08 && 0.923$\pm$0.010  \\\hline
8&&  $\checkmark$ &  $\checkmark$ &   & 27.45$\pm$1.26 && 0.915$\pm$0.021  \\\hline
9&& $\checkmark$  &   & $\checkmark$  & 28.28$\pm$1.17 && 0.933$\pm$0.012  \\\hline
10&&   & $\checkmark$  & $\checkmark$  & 28.66$\pm$1.05 && 0.943$\pm$0.007   \\\hline
11&$\checkmark$& $\checkmark$& $\checkmark$ & &28.52$\pm$1.38&&0.942$\pm$0.011 \\\hline
12&$\checkmark$& $\checkmark$&& $\checkmark$  &28.75$\pm$1.29&&0.946$\pm$0.008 \\\hline
13&& $\checkmark$&$\checkmark$&$\checkmark$  &29.43$\pm$1.47&& 0.951$\pm$0.003 \\\hline
\end{tabular}
\end{table}

\begin{table}[h]
  \centering
\caption{Quantitative results of our method on the IXI dataset. PSNR and SSIM are reported values are mean $\pm$ std.}\label{IXI_syn}
\begin{tabular}{lllllll}
\hline
&\multicolumn{3}{c}{Modalities} & \multicolumn{2}{c}{Results} \\
Number & T1 & T2 & PD &  PSNR($dB$)   &    & SSIM   \\\hline
1&$\checkmark$ &   &   &  34.00$\pm$1.59 &&  0.967$\pm$0.011    \\\hline
2&&  $\checkmark$ &   & 34.57$\pm$1.72 & & 0.971$\pm$0.012    \\\hline
3& &   & $\checkmark$  & 35.08$\pm$1.67 & & 0.974$\pm$0.008    \\\hline
4&$\checkmark$ & $\checkmark$  &   & 34.72$\pm$1.38 & & 0.972$\pm$0.007    \\\hline
5&$\checkmark$ &   & $\checkmark$  & 35.24$\pm$1.53 & & 0.976$\pm$0.011    \\\hline
6& &  $\checkmark$ & $\checkmark$  & 35.98$\pm$1.49 &  &0.978$\pm$0.008    \\\hline
\end{tabular}
\end{table}

\begin{table*}[]
\setlength \tabcolsep{3pt}
\caption{Quantitative comparison with SOTA methods on BraTS dataset. PSNR($dB$), SSIM, LPIPS($\times10^-2$), and FID are listed and reported values are mean $\pm$ std. The \textbf{boldface} indicates the top-performing model for each task.} 
\label{BraTS_sota}
\begin{tabular}{lllllllllllllllll}
   \hline
\multirow{2}{*}{} & \multicolumn{4}{c}{T1 → T1ce} & \multicolumn{4}{c}{T1ce→T1} & \multicolumn{4}{c}{T1,T2→T1ce} & \multicolumn{4}{c}{T1,FLAIR→T1ce} \\
 & \multicolumn{1}{c}{PSNR} & \multicolumn{1}{c}{SSIM} & \multicolumn{1}{c}{LPIPS} & \multicolumn{1}{c}{FID} & \multicolumn{1}{c}{PSNR} & \multicolumn{1}{c}{SSIM} & \multicolumn{1}{c}{LPIPS} & \multicolumn{1}{c}{FID} & \multicolumn{1}{c}{PSNR} & \multicolumn{1}{c}{SSIM} & \multicolumn{1}{c}{LPIPS} & \multicolumn{1}{c}{FID} & \multicolumn{1}{c}{PSNR} & \multicolumn{1}{c}{SSIM} & \multicolumn{1}{c}{LPIPS} & \multicolumn{1}{c}{FID} \\   \hline
pix2pix & 21.55 & 0.852 & 25.07 & 31.67 & 23.63 & 0.882 & 21.68 & 27.54 & 23.57 & 0.873 & 21.39 & 27.96 & 24.07 & 0.915 & 20.57 & 24.96 \\
 & $\pm$1.28 & $\pm$0.017 & $\pm$1.53 &  & $\pm$1.46 & $\pm$0.013 & $\pm$1.35 &  & $\pm$1.54 & $\pm$0.013 & $\pm$1.27 &  & $\pm$1.52 & $\pm$0.013 & $\pm$1.31 &  \\   \hline
pGAN & 22.26 & 0.857 & 23.76 & 29.56 & 24.19 & 0.887 & 20.17 & 25.37 & 24.23 & 0.875 & 20.37 & 25.72 & 24.45 & 0.919 & 19.16 & 24.03 \\
 & $\pm$1.19 & $\pm$0.013 & $\pm$1.39 &  & $\pm$1.29 & $\pm$0.016 & $\pm$1.28 &  & $\pm$1.27 & $\pm$0.015 & $\pm$1.42 &  & $\pm$1.37 & $\pm$0.016 & $\pm$1.07 &  \\   \hline
MM-GAN & 23.59 & 0.867 & 21.85 & 27.68 & 24.92 & 0.895 & 19.62 & 24.14 & 25.08 & 0.883 & 18.69 & 24.96 & 25.83 & 0.926 & 17.62 & 21.64 \\
 & $\pm$1.54 & $\pm$0.014 & $\pm$1.94 &  & $\pm$1.32 & $\pm$0.014 & $\pm$0.99 &  & $\pm$1.62 & $\pm$0.012 & $\pm$1.17 &  & $\pm$1.63 & $\pm$0.013 & $\pm$1.15 &  \\   \hline
Uni-GAN & 24.67 & 0.873 & 20.72 & 25.94 & 25.98 & 0.908 & 17.48 & 23.41 & 25.67 & 0.886 & 18.02 & 23.91 & 26.74 & 0.935 & 16.36 & 20.78 \\
 & $\pm$1.37 & $\pm$0.013 & $\pm$0.78 &  & $\pm$1.78 & $\pm$0.016 & $\pm$1.06 &  & $\pm$1.28 & $\pm$0.014 & $\pm$1.20 &  & $\pm$1.32 & $\pm$0.016 & $\pm$1.72 &  \\   \hline
ResVit & 24.32 & 0.871 & 21.38 & 26.51 & 25.51 & 0.903 & 18.08 & 23.97 & 25.77 & 0.889 & 17.73 & 23.57 & 26.58 & 0.933 & 16.94 & 21.96 \\
 & $\pm$1.62 & $\pm$0.009 & $\pm$1.14 &  & $\pm$1.38 & $\pm$0.015 & $\pm$1.25 &  & $\pm$1.64 & $\pm$0.015 & $\pm$0.92 &  & $\pm$1.48 & $\pm$0.014 & $\pm$1.47 &  \\   \hline
LDM & 23.25 & 0.866 & 22.19 & 28.53 & 25.08 & 0.896 & 18.96 & 24.03 & 24.74 & 0.880 & 19.11 & 24.76 & 25.19 & 0.922 & 18.54 & 22.59 \\
 & $\pm$1.08 & $\pm$0.010 & $\pm$1.42 &  & $\pm$1.34 & $\pm$0.010 & $\pm$1.53 &  & $\pm$1.42 & $\pm$0.012 & $\pm$1.17 &  & $\pm$1.52 & $\pm$0.012 & $\pm$1.26 &  \\   \hline
CoLa-Diff & 24.93 & 0.880 & 19.43 & 25.16 & 26.12 & 0.907 & 16.97 & 23.17 & 25.83 & 0.890 & 17.45 & 23.51 & 26.93 & 0.932 & 15.91 & 19.25 \\
 & $\pm$1.52 & $\pm$0.010 & $\pm$0.97 &  & $\pm$1.62 & $\pm$0.011 & $\pm$1.37 &  & $\pm$1.52 & $\pm$0.015 & $\pm$1.35 &  & $\pm$1.09 & $\pm$0.011 & $\pm$1.42 &  \\   \hline
Our method & \textbf{25.78} & \textbf{0.884} & \textbf{18.23} & \textbf{23.76} & \textbf{26.65} & \textbf{0.917} & \textbf{16.48} & \textbf{22.35} & \textbf{26.62} & \textbf{0.897} & \textbf{16.26} & \textbf{21.85} & \textbf{27.64} & \textbf{0.942} & \textbf{15.49} & \textbf{17.55} \\
 & \textbf{$\pm$1.63} & \textbf{$\pm$0.015} & \textbf{$\pm$1.06} & \textbf{} & \textbf{$\pm$1.30} & \textbf{$\pm$0.012} & \textbf{$\pm$1.42} & \textbf{} & \textbf{$\pm$1.37} & \textbf{$\pm$0.011} & \textbf{$\pm$1.28} & \textbf{} & \textbf{$\pm$1.33} & \textbf{$\pm$0.013} & \textbf{$\pm$1.08} & \textbf{}\\
\hline
\end{tabular}
\end{table*}

\begin{table*}[t]
    \centering
     \setlength \tabcolsep{3pt}
     \caption{Quantitative comparison with SOTA methods on IXI dataset. PSNR($dB$), SSIM, LPIPS($\times10^-2$), and FID are listed and reported values are mean $\pm$ std. The \textbf{boldface} indicates the top-performing model for each task.} \label{IXI_sota}
 \begin{tabular}{lllllllllllllllll}
 \hline
\multirow{2}{*}{} & \multicolumn{4}{c}{T1 → PD} & \multicolumn{4}{c}{PD→T1} & \multicolumn{4}{c}{T1,T2→PD} & \multicolumn{4}{c}{T1,PD→T2} \\
 & \multicolumn{1}{c}{PSNR} & \multicolumn{1}{c}{SSIM} & \multicolumn{1}{c}{LPIPS} & \multicolumn{1}{c}{FID} & \multicolumn{1}{c}{PSNR} & \multicolumn{1}{c}{SSIM} & \multicolumn{1}{c}{LPIPS} & \multicolumn{1}{c}{FID} & \multicolumn{1}{c}{PSNR} & \multicolumn{1}{c}{SSIM} & \multicolumn{1}{c}{LPIPS} & \multicolumn{1}{c}{FID} & \multicolumn{1}{c}{PSNR} & \multicolumn{1}{c}{SSIM} & \multicolumn{1}{c}{LPIPS} & \multicolumn{1}{c}{FID} \\\hline
pix2pix & 30.58 & 0.961 & 17.19 & 27.68 & 30.62 & 0.955 & 17.48 & 28.17 & 31.89 & 0.962 & 18.65 & 27.65 & 32.27 & 0.963 & 16.14 & 26.48 \\
 & $\pm$1.75 & $\pm$0.016 & $\pm$2.08 &  & $\pm$1.47 & $\pm$0.015 & $\pm$1.69 &  & $\pm$1.63 & $\pm$0.017 & $\pm$1.87 &  & $\pm$1.39 & $\pm$0.016 & $\pm$1.74 &  \\\hline
pGAN & 31.27 & 0.964 & 16.25 & 25.26 & 32.02 & 0.965 & 16.35 & 25.74 & 32.00 & 0.966 & 17.96 & 24.76 & 32.48 & 0.968 & 15.47 & 23.59 \\
 & $\pm$1.62 & $\pm$0.019 & $\pm$1.66 &  & $\pm$1.38 & $\pm$0.013 & $\pm$1.48 &  & $\pm$1.28 & $\pm$0.016 & $\pm$1.38 &  & $\pm$1.62 & $\pm$0.016 & $\pm$1.95 &  \\\hline
MM-GAN & 32.19 & 0.966 & 15.63 & 22.14 & 32.64 & 0.968 & 15.79 & 23.74 & 33.02 & 0.969 & 16.38 & 23.16 & 33.22 & 0.967 & 14.34 & 21.65 \\
 & $\pm$1.53 & $\pm$0.013 & $\pm$1.48 &  & $\pm$1.53 & $\pm$0.016 & $\pm$1.57 &  & $\pm$1.63 & $\pm$0.014 & $\pm$1.56 &  & $\pm$1.38 & $\pm$0.014 & $\pm$1.68 &  \\\hline
Uni-GAN & 33.54 & 0.973 & 14.25 & 18.96 & 34.68 & 0.974 & 13.08 & 17.95 & 34.23 & 0.973 & 13.79 & 18.38 & 35.81 & 0.973 & 12.18 & 16.74 \\
 & $\pm$1.96 & $\pm$0.015 & $\pm$1.62 &  & $\pm$1.29 & $\pm$0.012 & $\pm$1.43 &  & $\pm$1.57 & $\pm$0.013 & $\pm$1.47 &  & $\pm$1.48 & $\pm$0.015 & $\pm$1.37 &  \\\hline
ResVit & 33.26 & 0.972 & 15.37 & 20.79 & 34.51 & 0.972 & 13.57 & 19.42 & 33.95 & 0.975 & 14.53 & 19.52 & 35.63 & 0.977 & 12.67 & 17.32 \\
 & $\pm$1.73 & $\pm$0.013 & $\pm$1.44 &  & $\pm$1.42 & $\pm$0.010 & $\pm$1.33 &  & $\pm$1.73 & $\pm$0.013 & $\pm$1.38 &  & $\pm$1.54 & $\pm$0.013 & $\pm$1.22 &  \\\hline
LDM & 32.88 & 0.971 & 13.98 & 18.34 & 33.53 & 0.970 & 14.92 & 21.95 & 33.43 & 0.973 & 15.72 & 22.51 & 33.99 & 0.972 & 13.79 & 20.78 \\
 & $\pm$1.48 & $\pm$0.012 & $\pm$1.65 &  & $\pm$1.34 & $\pm$0.009 & $\pm$1.76 &  & $\pm$1.56 & $\pm$0.015 & $\pm$1.19 &  & $\pm$1.18 & $\pm$0.016 & $\pm$1.43 &  \\\hline
CoLa-Diff & 33.53 & 0.972 & 14.52 & 19.14 & 34.27 & 0.976 & 14.08 & 18.42 & 33.82 & 0.973 & 14.92 & 20.47 & 35.29 & 0.975 & 13.09 & 18.53 \\
 & $\pm$1.42 & $\pm$0.015 & $\pm$1.59 &  & $\pm$1.08 & $\pm$0.012 & $\pm$1.37 &  & $\pm$1.27 & $\pm$0.009 & $\pm$1.32 &  & $\pm$1.52 & $\pm$0.013 & $\pm$1.49 &  \\\hline
Our method & \textbf{34.17} & \textbf{0.973} & \textbf{13.56} & \textbf{17.93} & \textbf{35.23} & \textbf{0.981} & \textbf{12.16} & \textbf{15.38} & \textbf{34.72} & \textbf{0.977} & \textbf{13.17} & \textbf{17.46} & \textbf{36.24} & \textbf{0.982} & \textbf{11.65} & \textbf{15.89} \\
 & \textbf{$\pm$1.58} & \textbf{$\pm$0.014} & \textbf{$\pm$1.37} & \textbf{} & \textbf{$\pm$1.27} & \textbf{$\pm$0.014} & \textbf{$\pm$1.58} & \textbf{} & \textbf{$\pm$1.46} & \textbf{$\pm$0.010} & \textbf{$\pm$1.49} & \textbf{} & \textbf{$\pm$1.43} & \textbf{$\pm$0.004} & \textbf{$\pm$1.42} & \textbf{} \\ \hline
\end{tabular}
\end{table*}

\subsection{Synthesis results of task-specific comparison with state-of-the-art }  \label{task}

To evaluate the synthesis performance in task-specific (one-to-one and many-to-one), task-specific FgC2F-UDiff was compared with the other seven SOTA methods (pix2pix, pGAN, MM-GAN, Uni-GAN, ResVit, LDM, and CoLa-Diff) on the BraTS and IXI datasets. The quantitative results are shown in Table.\ref{BraTS_sota} and Table.\ref{IXI_sota}, which indicated that our proposed FgC2F-UDiff achieves the best performance in both task-specific (one-to-one and many-to-one) in terms of PSNR, SSIM, LPIPS, and FID metrics ($p < 0.05$). Specifically, on the BraTS dataset, one-to-one tasks of T1 $\rightarrow$ T1ce; T1ce $\rightarrow$ T1, many-to-one tasks of T1,T2 $\rightarrow$ T1ce; and T1, FLAIR $\rightarrow$ T1ce were considered, as shown in Table.\ref{BraTS_sota}.  In the one-to-one task of T1 $\rightarrow$ T1ce, our method outperforms pix2pix, pGAN, MM-GAN, Uni-GAN, ResVit, LDM, and CoLa-Diff by a margin of 4.23$dB$, 3.52$dB$, 2.19$dB$, 1.11$dB$, 1.46$dB$, 2.53$dB$, and 0.85$dB$, respectively.  On the IXI dataset, one-to-one tasks of T1 $\rightarrow$ PD; PD $\rightarrow$ T1, many-to-one tasks of T1,T2 $\rightarrow$ PD; and T1, PD $\rightarrow$ T2 were considered, as shown in Table.\ref{IXI_sota}. Our proposed FgC2F-UDiff achieves the best performance in both one-to-one tasks and many-to-one tasks in terms of PSNR, SSIM, LPIPS, and FID metrics ($p < 0.05$). Specifically, in the one-to-one task of T1 $\rightarrow$ PD, our method outperforms pix2pix, pGAN, MM-GAN, Uni-GAN, ResVit, LDM and CoLa-Diff by a margin of 3.59$dB$, 2.9$dB$, 1.98$dB$, 0.63$dB$, 0.91$dB$, 1.29$dB$,  and 0.64$dB$, respectively. All these quantitative results prove that our FgC2F-UDiff is superior to other methods in the medical image synthesis task.

\begin{figure*}[tp]
\center
\includegraphics[width=1\textwidth]{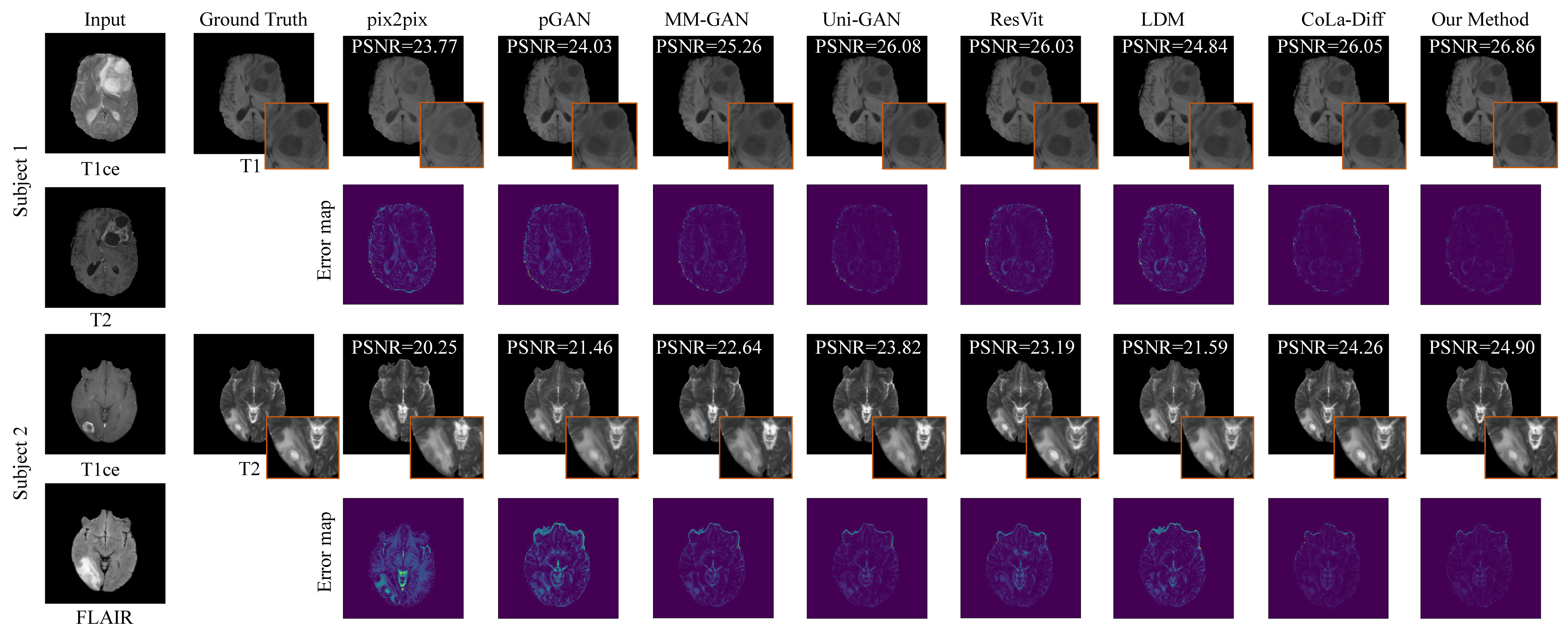}
\caption{Illustrative instances of synthetic images were demonstrated on the BraTS dataset for two representative tasks. Synthesized images from all competing methods are shown along with the source and reference target images. Partial enlargement and error plots can more intuitively observe the differences between the synthesized image and the ground truth, thereby reflecting the quality of the synthesis.} \label{Sota_BraTS}
\end{figure*}

\begin{figure*}[tp]
\center
\includegraphics[width=1\textwidth]{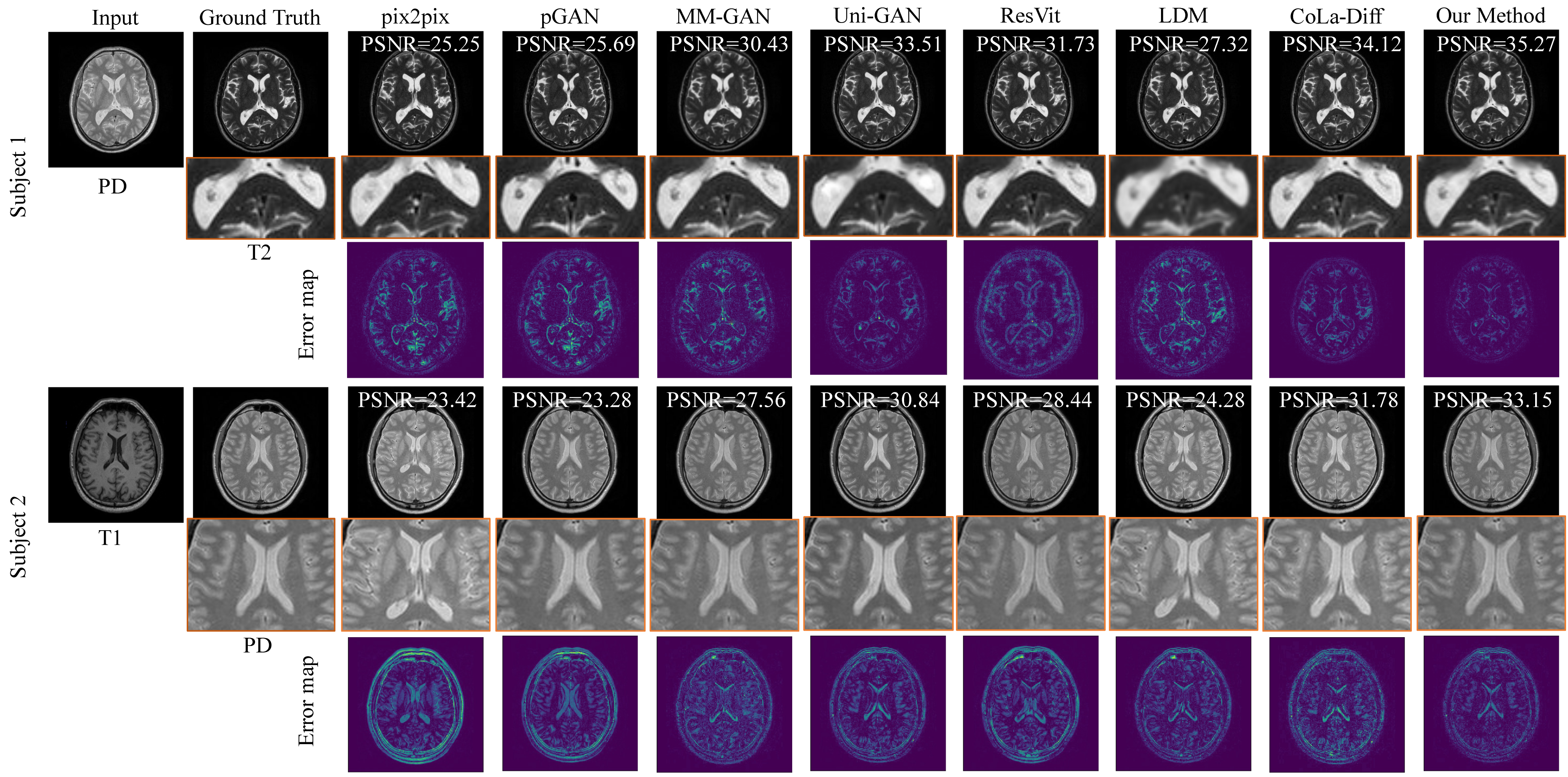}
\caption{Illustrative instances of synthetic images were demonstrated on the IXI dataset for two representative synthesis tasks. Synthesized images from all competing methods are shown along with the source and reference target images. Partial enlargement and error plots can more intuitively observe the differences between the synthesized image and the ground truth, thereby reflecting the quality of the synthesis.} \label{Sota_IXI}
\end{figure*}

The visualized comparison results are shown in Fig.\ref{Sota_BraTS} and Fig.\ref{Sota_IXI}, which indicated that FgC2F-UDiff gains the best-synthesized performance compared with other SOTA methods on both datasets. Specifically, on the BraTS dataset, we selected two representative synthesis tasks: (1) T1ce, T2 $\rightarrow$ T1 and (2) T1ce, FLAIR $\rightarrow$ T2. The synthesized results for these tasks are displayed in Fig.\ref{Sota_BraTS}. To facilitate a more intuitive examination, we zoomed in on key regions within the synthesized results, which exhibit relatively conspicuous differences from the ground truth. Upon close observation of Subject 2, the disparities in synthesis primarily manifested in the tumor region. In comparison to the SOTA approaches, our proposed method closely approximated the ground truth.  Other more advanced methods (i.e., LDM model and ResVit) have also achieved good synthesis results. Methods based on GANs (i.e., Uni-GAN, MM-GAN, and pGAN) exhibited some errors in fine details, while pix2pix nearly failed to capture the tumor-enhanced area. On the IXI dataset, we selected two representative synthesis tasks: (1) PD $\rightarrow$ T2 and (2) T1 $\rightarrow$ PD, and the synthesized results are shown in Fig.\ref{Sota_IXI}. We magnified pivotal regions within the synthesized results to facilitate a more intuitive assessment that displayed discernible deviations from the corresponding ground truth. The results show that the synthesis results generated by our proposed FgC2F-UDiff model stand out notably. They exhibit superior quality are characterized by lower noise and retain better detail and structural information. The consistency observed between the quantitative and qualitative findings further demonstrates the superior synthesis performance of our proposed FgC2F-UDiff model. Compared to other state-of-the-art (SOTA) methods, FgC2F-UDiff consistently generates higher-quality results characterized by capturing fine details and preserving certain structural information with lower noise and clearer texture details, edges, and shapes.

\subsection{Synthesis results of unified model comparison with state-of-the-art }  \label{uni}

To evaluate the synthesis performance in unified synthesis models, unified FgC2F-UDiff was compared with the MM-GAN, ResVit, and Uni-GAN on many-to-one tasks of BraTS. Task-specific models are trained and tested to perform a single synthesis task to improve performance, but a separate model has to be built for each task. So, we demonstrate FgC2F-UDiff in learning unified synthesis models for multi-modality MRI. The quantitative results are shown in Table.\ref{Brats_uni} and Table.\ref{IXI_uni}, which indicated that our proposed unified FgC2F-UDiff achieves the best performance on a many-to-one task in terms of PSNR, SSIM, LPIPS, and FID metrics ($p < 0.05$). Specifically, on the BraTS dataset, many-to-one tasks of T1,T2 $\rightarrow$ T1ce and T1, FLAIR $\rightarrow$ T1ce were considered. As shown in Table.\ref{Brats_uni}, our proposed FgC2F-UDiff achieves the best performance in many-to-one tasks in terms of PSNR, SSIM, LPIPS, and FID metrics ($p < 0.05$). On the many-to-one task of T1, T2 $\rightarrow$ T1ce, our method outperforms MM-GAN, Uni-GAN, and ResVit by a margin of 1.55$dB$, 0.77$dB$, and 0.91$dB$, respectively. On the IXI dataset, many-to-one tasks of T1,T2 $\rightarrow$ PD and T1, PD $\rightarrow$ T2 were considered. As shown in Table.\ref{IXI_uni}, our proposed FgC2F-UDiff achieves the best performance in many-to-one tasks in terms of PSNR, SSIM, LPIPS, and FID metrics ($p < 0.05$). Specifically, in the many-to-one task of T1, T2 $\rightarrow$ PD, our method outperforms MM-GAN, Uni-GAN, and ResVit by a margin of 1.01$dB$, 0.60$dB$, and 0.93$dB$, respectively.

\begin{table}[t]
    \centering
     \setlength \tabcolsep{2pt}
     \caption{Quantitative comparison with SOTA methods. PSNR($dB$), SSIM, LPIPS($\times10^-2$), and FID are listed and reported values are mean $\pm$ std. The \textbf{boldface} indicates the top-performing model for each task.} \label{Brats_uni}
    \begin{tabular}{lllllllllll} 
    \hline
      \multirow{2}{*}{} & \multicolumn{4}{c}{T1,T2 $\rightarrow$ T1ce} & \multicolumn{4}{c}{T1, FLAIR $\rightarrow$ T1ce } \\ \hline
          & PSNR & SSIM & LPIPS&FID&PSNR & SSIM &LPIPS&FID \\ \hline
\multirow{2}{*}{MM-GAN} &24.79 &0.880 &19.58 &24.56 &26.67&0.923&18.46&21.93 \\ 
 &$\pm$1.43 &$\pm$0.011 &$\pm$1.38&& $\pm$1.38 &$\pm$0.024 &$\pm$1.42 & \\ \hline
\multirow{2}{*}{Uni-GAN}&25.57 &0.886 &18.13&21.77&26.54 &0.930&16.97&20.07 \\ 
 &$\pm$1.52 &$\pm$0.014&$\pm$1.28& &$\pm$1.23 &$\pm$0.028 &$\pm$1.08 & \\ \hline    
 \multirow{2}{*}{ResVit} &25.43&0.885&18.49&22.34&26.34 &0.929 &17.28&20.76 \\ 
   &$\pm$1.29 &$\pm$0.009&$\pm$1.73& &$\pm$1.45 &$\pm$0.031 &$\pm$1.19 &\\ \hline       
 \multirow{2}{*}{Our method} &\textbf{26.34} &\textbf{0.894} &\textbf{17.45}&\textbf{20.85}&\textbf{27.19} &\textbf{0.937} & \textbf{16.14} & \textbf{18.74} \\ 
   &\textbf{$\pm$1.07} &\textbf{$\pm$0.013} &\textbf{$\pm$1.32} && \textbf{$\pm$1.16} & \textbf{$\pm$0.017} &\textbf{$\pm$1.54} &\\ \hline
    \end{tabular}
\end{table}

\begin{table}[t]
    \centering
     \setlength \tabcolsep{2pt}
     \caption{Quantitative comparison with SOTA methods. PSNR($dB$), SSIM, LPIPS($\times10^-2$), and FID are listed and reported values are mean $\pm$ std. The \textbf{boldface} indicates the top-performing model for each task.} \label{IXI_uni}
    \begin{tabular}{lllllllllll} 
    \hline
      \multirow{2}{*}{} & \multicolumn{4}{c}{T1,T2 $\rightarrow$ PD} & \multicolumn{4}{c}{T1, PD $\rightarrow$ T2 }  \\ \hline
                & PSNR & SSIM & LPIPS&FID&PSNR & SSIM &LPIPS&FID \\ \hline
\multirow{2}{*}{MM-GAN} &31.48 &0.958&17.25&21.72&31.98 &0.963 &15.93&20.63 \\ 
  &$\pm$1.37 &$\pm$0.007&$\pm$1.55& &$\pm$1.53 &$\pm$0.011 &$\pm$1.67& \\ \hline
\multirow{2}{*}{Uni-GAN}&31.89&0.968&16.28&19.72 &32.90 &0.971 &14.97&17.44 \\ 
&$\pm$1.77 &$\pm$0.006&$\pm$1.36&&$\pm$1.27 &$\pm$0.008 &$\pm$1.26 &\\ \hline
\multirow{2}{*}{ResVit} &31.56 &0.963&16.93&20.35 &32.17 &0.971&15.42 &18.35 \\ 
  &$\pm$1.53 &$\pm$0.007&1.28& &$\pm$1.48 &$\pm$0.005 &$\pm$1.22 & \\ \hline
\multirow{2}{*}{Our method}& \textbf{32.49} & \textbf{0.971}& \textbf{15.73}& \textbf{18.13} & \textbf{33.77} & \textbf{0.973} & \textbf{14.52} & \textbf{16.38} \\ 
 & \textbf{$\pm$1.69} & \textbf{$\pm$0.005}& \textbf{$\pm$1.43}& & \textbf{$\pm$1.61} & \textbf{$\pm$0.007} & \textbf{$\pm$1.21} & \\ \hline
    \end{tabular}
\end{table}

\begin{figure}[h]
\center
\includegraphics[width=0.5\textwidth]{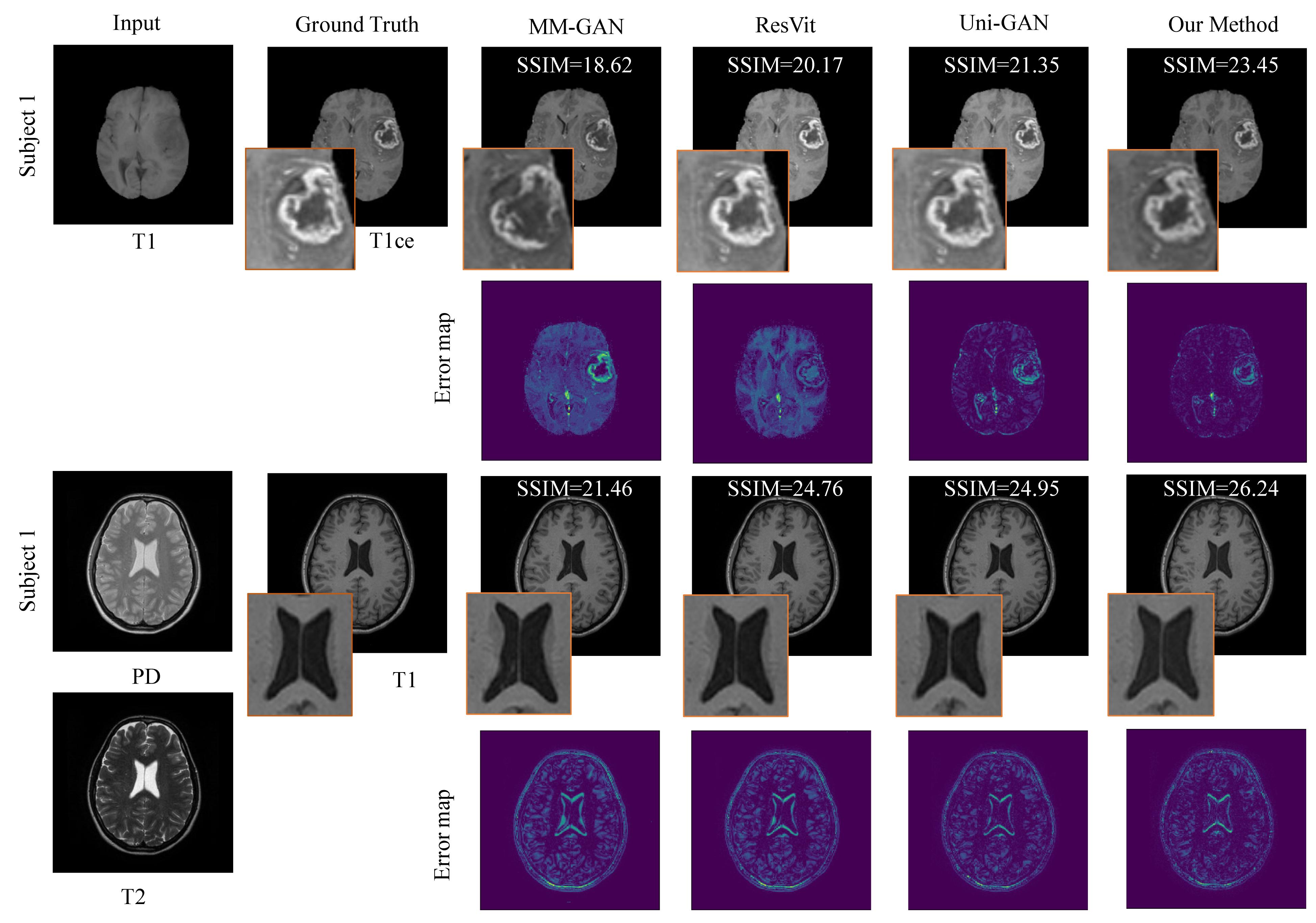}
\caption{Illustrative instances of synthetic images were demonstrated on the two datasets in learning unified synthesis models. Synthesized images from all competing methods are shown along with the source and reference target images.} \label{Uni}
\end{figure}

The visualized comparison results are shown in Fig.\ref{Uni}, we report two representative synthesis tasks of T1 $\rightarrow$ T1ce on BraTS and PD, T2 $\rightarrow$ T1 on IXI, which indicated that FgC2F-UDiff gains the best-synthesized performance compared with other SOTA methods on both datasets. Our method demonstrates the synthesis of target images characterized by lower noise and clearer texture details, edges, and shapes when compared to baseline models. These results indicate the proficient consolidation of models for diverse source-target configurations by the unified FgC2F-UDiff model. Additionally, the consistency observed between quantitative and qualitative research findings further validates the superior synthesis performance of our proposed FgC2F-UDiff model in learning unified synthesis models. 

\subsection{Ablation studies show improvements of innovation} \label{ablation} 

To validate the contributions of CUN, FCS, and SHM for cross-modality synthesis, we performed the comparison among our FgC2F-UDiff, the FgC2F-UDiff without CUN, the FgC2F-UDiff without FCS, and the FgC2F-UDiff without SHM. The structure of these ablation studies. Specifically, 1) To verify the contribution of our proposed CUN, we remove the module of frequency-guided conditions, leaving only the original diffusion model for synthesis tasks  (Ours w/o CUN), as shown in Fig.\ref{ablation}(b). All variants have the same network structure except for the CUN.  2) To further examine the contribution of learning non-linear mapping guided by the proposed FCS, we designed two ablation (as shown in Fig.\ref{ablation}(c)): one only utilizes all available modalities to obtain modality-class features (Ours w/o $LF or HF$), while the other solely employs frequency information for generating frequency-class guided features (Ours w/o ${Z_{T}^{m}}$). 3) To validate the contribution of SHM, we conducted the following experiments: (a) Our approach without CL (Ours w/o CL). We employ CL during training to facilitate network progression from easy to challenging tasks, thereby expediting both learning and the quality of synthesized samples. (b) Our approach without frequency-guided conditions (Ours w/o CUN). We leverage the frequency to guide the diffusion models to fit temporal regularities inherent in diffusion, thereby accelerating the coarse-to-fine synthesis.  (c) Our approach without dynamically selecting frequency-information (Ours w/o dsf). To maintain fairness in the experiment, we use high-frequency and low-frequency information from the same available image as conditions. We utilize a dynamic strategy to retain specific features, elevating the synthesis quality of complex regions within images. 

\begin{figure}[t]
\center
\includegraphics[width=0.4\textwidth]{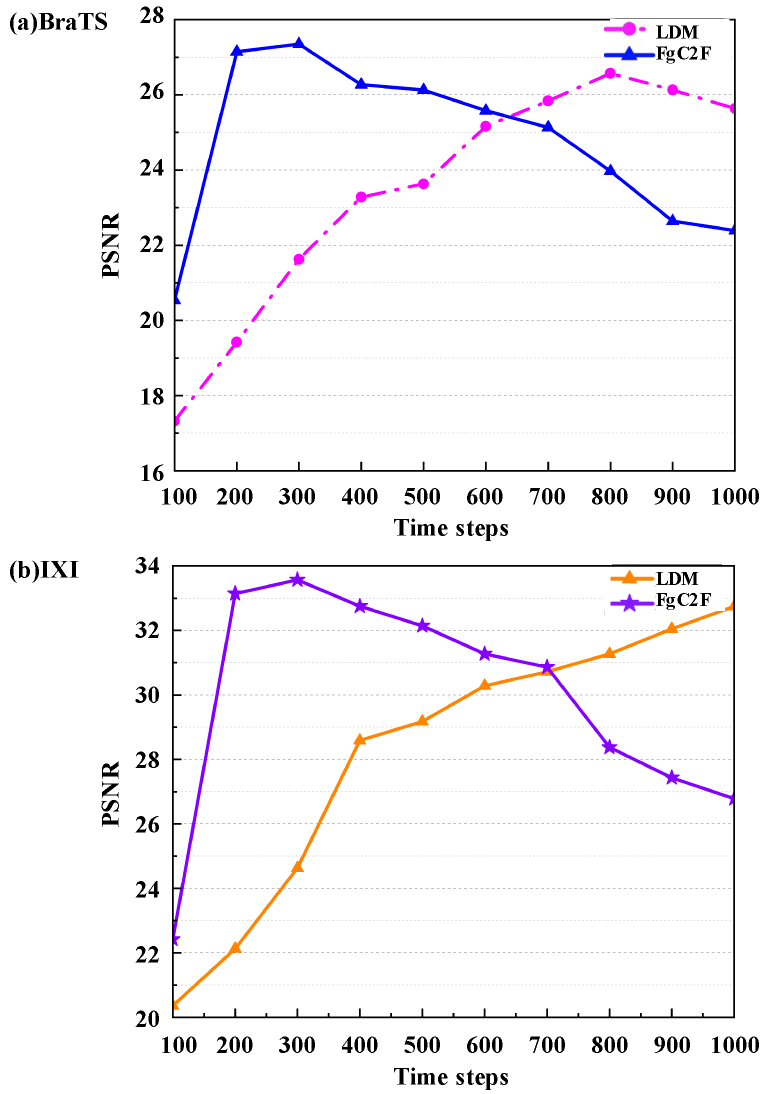}
\caption{The experimental results of our method compared with LDM at different time steps. The upper plot shows  PSNR values of LDM and FgC2F-UDiff under different time steps on BraTS. The lower plot shows PSNR values of LDM and FgC2F-UDiff at different time steps on IXI.} \label{timestep}
\end{figure}

The quantitative analysis results of the ablation study are presented in Table.\ref{Brats_ablation}, which demonstrated that every part of FgC2F-UDiff contributes to the cross-modality synthesis. We report the average numerical results of each method across 14 input scenarios on the BraTS dataset. Especially for the CUN and $Z_{T}^{m}$ all bring the cross-modality synthesis's clear performance gain. Specifically, for T1,T2 $\rightarrow$ T1ce, the PSNR decreased from 26.34 $\pm$ 1.07 to 20.58 $\pm$1.53 when moving the CUN and decreased from 26.34 $\pm$ 1.07 to 22.64 $\pm$ 0.94 when moving the $Z_{T}^{m}$. 

\begin{table}[t]
    \centering
    \setlength \tabcolsep{2pt}
     \caption{ Quantitative comparison of ablation study. PSNR($dB$), SSIM, LPIPS($\times10^-2$), and FID are listed and reported values are mean $\pm$ std. The \textbf{boldface} indicates the top-performing model for each task.} \label{Brats_ablation}
    \begin{tabular}{lllllllllll} 
    \hline
      \multirow{2}{*}{} & \multicolumn{4}{c}{T1,T2 $\rightarrow$ T1ce} & \multicolumn{4}{c}{T1, FLAIR $\rightarrow$ T1ce } \\ \hline
          & PSNR & SSIM & LPIPS&FID&PSNR & SSIM &LPIPS&FID \\ \hline
\multirow{2}{*}{W/O CUN} &20.58&0.873&22.42&31.17& 21.21 &0.907 &21.04 &28.86 \\ 
&$\pm$1.53 &$\pm$0.009&$\pm$1.85& &$\pm$1.34 &$\pm$0.011 &$\pm$1.46 & \\ \hline
\multirow{2}{*}{W/O ${Z_{T}^{m}}$} &22.64&0.880&18.28&22.18&23.57&0.919 &19.43 &25.14 \\ 
&$\pm$0.94 &$\pm$0.013&$\pm$1.37& &$\pm$1.27 &$\pm$0.009 &$\pm$1.53 &\\ \hline
W/O $LF$ &25.35 &0.889&18.28&22.14 &26.11&0.934&17.25&20.97\\ 
or HF &$\pm$1.27 &$\pm$0.011&$\pm$1.36& &$\pm$1.19 &$\pm$0.014 &$\pm$1.18 & \\ \hline
\multirow{2}{*}{ W/O CL} &24.52&0.883&18.02&23.86&25.15 &0.929 & 18.63& 22.42 \\ 
&$\pm$1.17 &$\pm$0.008&$\pm$1.52& &$\pm$1.45 &$\pm$0.012 & $\pm$1.08&  \\ \hline
\multirow{2}{*}{ W/O dsf} & 25.87&0.890&17.83&21.23 &26.63 &0.932 &16.87 &20.35 \\ 
& $\pm$1.32 &$\pm$0.013&$\pm$1.22& &$\pm$1.16 &$\pm$0.017 &$\pm$1.06 &\\ \hline
  \multirow{2}{*}{Our method} &\textbf{26.34} &\textbf{0.894} &\textbf{17.45}&\textbf{20.85}&\textbf{27.19} &\textbf{0.937} & \textbf{16.14} & \textbf{18.74} \\ 
   &\textbf{$\pm$1.07} &\textbf{$\pm$0.013} &\textbf{$\pm$1.32} && \textbf{$\pm$1.16} & \textbf{$\pm$0.017} &\textbf{$\pm$1.54} &\\ \hline
    \end{tabular}
\end{table}

\subsection{Analysis of SHM} \label{SHM1}

To verify that our proposed acceleration mechanism can reduce the denoising time steps and inference time of our method without reducing the synthesis quality, we analyzed the impact of the SHM on our FgC2F-UDiff. In Fig.\ref{timestep}, a comparative analysis of our approach, which integrates the LDM with SHM, reveals notable advancements in network training efficiency and a reduction in denoising steps. Specifically, when examining the BraTS dataset, it becomes evident that utilizing our method with a $T$ of 200 or 300 surpasses the performance of the standalone LDM with a $T$ set of 800. Likewise, when evaluating the IXI dataset, our model exhibits superior results with $T$ values of 200 or 300 compared to the LDM with $T$=1000. Moreover, the decline in performance metrics beyond 300-time steps appears to be due to 'over-diffusion' \cite{song2020denoising}, where the model continues to apply noise beyond the optimal range for image synthesis, resulting in unnecessary alterations that degrade image quality. Our experimental results clearly demonstrate that our method, combined with the proposed SHM, significantly accelerates the sampling speed, effectively addressing the sluggishness typically associated with conventional diffusion models. In addition, this discrepancy in performance can be attributed to several factors between the different datasets. The BraTS dataset boasts a greater number of modalities, thereby offering a richer array of diverse features for the model to leverage. Moreover, the increased sample size in BraTS contributes to heightened sample diversity, further enhancing the performance of FgC2F-UDiff. 

Furthermore, our empirical analysis conclusively demonstrates that our proposed method significantly reduces inference times compared to the LDM. Utilizing 200 reverse diffusion steps, our method achieves an inference time of approximately 1.5 to 2 seconds per image in BraTS. In stark contrast, LDM, requiring 800 timesteps, has an inference time ranging from 3.0 to 4.5 seconds per image under identical hardware conditions. This substantial reduction in inference time highlights the efficiency and practicality of our approach in real-world applications where rapid image synthesis is critical.

\section{Conclusion} 
This paper has presented a unified network for multi-modality missing MRI synthesis using a Frequency-guided and Coarse-to-fine Unified Diffusion Model (FgC2F-UDiff) from multiple inputs and outputs. CUN network has been introduced to leverage iterative denoising properties of the diffusion model to improve the fidelity of synthesizing images. In addition, an FCS strategy has been designed to utilize the frequency information to guide coarse-to-fine synthesis.
The SHM further accelerates the diffusion process by intelligently integrating specific mechanisms, enhancing the efficiency and practicality of FgC2F-UDiff. Extensive experimental evaluations on two medical image synthesis datasets validate the effectiveness of our approach. This study provides a new perspective for addressing the missing modality issue in current technologies.

\bibliographystyle{IEEEtran}
\bibliography{mybibfile}

\end{document}